\documentclass[12pt]{article}
\usepackage{amssymb,graphicx}

\makeatletter
\renewcommand\section{\@startsection {section}{1}{\z@}%
                                 {-3.5ex \@plus -1ex \@minus -.2ex}
                                   {2.3ex \@plus.2ex}%
                                   {\normalfont\large\bfseries}}
\renewcommand\subsection{\@startsection{subsection}{2}{\z@}%
                                   {-3.25ex\@plus -1ex \@minus -.2ex}%
                                     {1.5ex \@plus .2ex}%
                                     {\normalfont\bfseries}}
\renewcommand\subsubsection{\@startsection{subsubsection}{3}{\z@}%
                                   {-3.25ex\@plus -1ex \@minus -.2ex}%
                                     {1.5ex \@plus .2ex}%
                                     {\normalfont\itshape}}
\makeatother

\newcommand{\Letter}{
    \setlength{\textwidth}{7in}
    \setlength{\textheight}{9.5in}
    \hoffset=-0.75in
    \voffset=-1.15in }

\Letter

\newcommand{\ZZ}{\mathbb{Z}}

\newcommand{\RR}{\mathbb{R}}

\newcommand{\gsim}{ \lower .75ex \hbox{$\sim$} \llap{\raise .27ex \hbox{$>$}} }
\newcommand{\lsim}{ \lower .75ex \hbox{$\sim$} \llap{\raise .27ex \hbox{$<$}} }

\def\beq{\begin{equation}}
\def\eeq{\end{equation}}

\def\D3{\overline{\rm D3}}

\begin{document}

\begin{titlepage}
\setcounter{page}{1} \baselineskip=15.5pt \thispagestyle{empty}
\begin{flushright}
\parbox[t]{2in}{
COLO-HEP-524\\
MAD-TH-06-14\\
PUPT-2224\\
{\tt hep-th/yymmnnn}}
\end{flushright}

\vfil


\begin{center}
{\LARGE D3-brane Vacua in Stabilized Compactifications}
\end{center}
\bigskip

\begin{center}
%
{Oliver DeWolfe,$^1$ Liam McAllister,$^2$ Gary Shiu,$^3$ and Bret
Underwood$^3$}
\end{center}

\begin{center}
${}^1$\textit{Department of Physics, 390 UCB,
     University of Colorado,
     Boulder, CO 80309, USA}
\vskip 3pt
${}^2$\textit{Department of Physics,
     Princeton University,
     Princeton, NJ 08544, USA}
\vskip 3pt
${}^3$\textit{Department of Physics,
     University of Wisconsin,
     Madison, WI 53706, USA}
\end{center}
\bigskip \bigskip \bigskip \bigskip

\begin{center}
{\bf
Abstract}
\end{center}

\noindent D3-branes feel no force in no-scale flux
compactifications of type IIB string theory, but the
nonperturbative effects required to stabilize the K\"ahler moduli
break the no-scale structure and generate a potential for D3-brane
motion, confining the branes to certain loci.
D3-branes away from these loci break supersymmetry spontaneously,
by an F-term.  We present the general conditions for
supersymmetric D3-brane vacua in models with a single K\"ahler modulus,
then explicitly calculate these vacua for
D3-branes moving on the tip of the warped deformed conifold.
We find both continuous moduli spaces and isolated vacua.
In addition, we show that $\D3$-branes
and D3-branes are localized to the same
regions by the nonperturbative potential,
avoiding a potential obstacle to brane inflation.
We apply these results
to determine whether angular motion of a brane in a throat could
play an important role in inflation, and find that any
inflation along the angular directions is short-lived because the
field space is very small.

\vfil
\begin{flushleft}
\today
\end{flushleft}

\end{titlepage}

\newpage



\tableofcontents

\newpage

\section{Introduction}
\label{IntroSec}

D-branes have
proved to be an extremely useful tool in constructing
models of particle physics and cosmology (for recent reviews,
see {\it{e.g.}} \cite{reviews,reviewscosmo}).
In type IIB string
theory, D3-branes play a particularly distinctive role.
A D3-brane that fills spacetime is pointlike in the internal space, and so the
configuration space of the D3-brane is the entire compactification
manifold.  Moreover, in a no-scale flux compactification \cite{GKP,Noscalefluxpapers},
D3-branes feel no potential (to leading order in $\alpha'$ and $g_s$), and thus
the D3-brane moduli space can in fact be identified with the
internal space.  In contrast, higher-dimensional branes wrapping
various cycles have more complicated configuration spaces, and
also generically receive potentials from the flux background.
As the most mobile spacetime-filling D-branes,
D3-branes provide a key ingredient in models of brane inflation \cite{DvaliTye,otherbraneinflation}.
The idea of brane inflation
is  to exploit the mobility of D3-branes
and to look for a weak force
that slightly lifts the D3-brane moduli space, producing a relatively
flat potential on the internal manifold.
The inflaton could then be identified as the D3-brane position.

However, in a compactification in which the K\"ahler moduli are
stabilized by nonperturbative effects, as
described in \cite{KKLT},
the no-scale structure is broken and
D3-branes no longer enjoy a no-force condition. A D3-brane at a generic point then breaks
supersymmetry spontaneously, by an F-term associated with the
nonperturbative superpotential.  The D3-brane feels a force from
nonperturbative effects that confines it to certain restricted
loci where the F-term vanishes and supersymmetry is restored.
As we shall see, there are generically enough constraints to reduce the space of D3-brane vacua to
a set of isolated points, but many
concrete models preserve isometries,
so that associated moduli spaces are possible as well.
The effects of the moduli-stabilizing F-term on D-brane inflation have recently been
explored in \cite{KKLMMT,Baumann,Uplifting}; given the ubiquity of D3-branes in
the construction of Standard-like models \cite{reviews}, finding the D3-branes vacua in stabilized compactifications
is highly relevant for particle physics considerations as well.

In this paper we study the general
conditions for supersymmetric
D3-brane vacua in moduli-stabilized type IIB flux
compactifications with a single K\"ahler modulus.
These conditions depend
on the compactification geometry and on the embedding of
the moduli-stabilizing branes.  Furthermore, we study a number of
explicit examples in the warped deformed conifold \cite{KS},
characterized by particular supersymmetric embeddings of the D7-branes (or Euclidean D3-branes).
In any given compact model, one or more of the possible supersymmetric embeddings  might generate a nonperturbative superpotential; in this paper we assume such a superpotential is generated and study the consequences.  We consider two broad classes of embeddings in the
local throat model, and obtain
moduli spaces of real dimension zero, one, and two.

Brane/antibrane inflation relies on the final annihilation of the pair as an exit mechanism.  Correspondingly, it is a natural question whether the $\D3$-branes are confined to the same loci by the nonperturbative effects as the D3-branes --
if it were otherwise, a new potential could interfere with the end of inflation.  We show that D3-branes and $\D3$-branes
are in fact localized to the same loci on the tip of the warped deformed conifold.

The angular directions in the warped deformed conifold are protected by isometries, making them potentially attractive as inflationary directions.  Moreover, the warping isolates the bottom of the throat from the bulk of the geometry and raises the possibility of flattening out the brane Coulomb potential.
We investigate this possibility of D3-brane inflation corresponding to motion along an approximately preserved moduli space at the
tip of the warped deformed conifold.  We find that although the warp factor naively flattens the potential, it also compresses the field space, and consequently angular inflation is not a viable mechanism.

The organization of this paper is as follows.  In \S\ref{ReviewSec} we briefly
review warped flux compactifications of type IIB string theory. In
\S\ref{SUSYVacSec} we determine the equations for
supersymmetric D3-brane vacua in the presence of moduli stabilization
forces in all generality, and in \S\ref{ConifoldVacSec} we apply
these results to a number of explicit examples on the $S^3$ at the
tip of the warped deformed conifold.  In \S\ref{AntibraneSec} we
study the analogous problem for $\D3$-branes and find that
$\D3$-branes and D3-branes are confined to the same loci on the
tip.  In \S\ref{InflationSec},
we enumerate all possible forces on branes at the tip of the throat, and
examine whether angular motion of a brane/antibrane system there can significantly inflate, obtaining a negative result.
We conclude in \S\ref{ConclusionsSec}. Some details are relegated to the appendices.

\section{Review of type IIB compactifications}
\label{ReviewSec}
\setcounter{equation}{0}

We consider a warped compactification of type IIB string theory
down to four dimensions, with metric ansatz \begin{eqnarray}
\label{metric} ds^2 = e^{2A(y)} e^{-6u(x)} g_{\mu\nu} dx^\mu
dx^\nu + e^{2u(x)} e^{-2A(y)} \tilde{g}_{mn} dy^m dy^n \,,
\end{eqnarray}
where $e^{A(y)}$ is the warp factor
and $e^{2u(x)}$ is the Weyl rescaling
required to decouple the overall volume mode from the
four-dimensional graviton.
We take the ``unwarped" metric $\tilde{g}_{mn}$ on the compact space to be
Calabi-Yau;\footnote{In more generality, when 7-branes are present, the unwarped
metric/dilaton system corresponds to an F-theory compactification
on a Calabi-Yau fourfold.}
we will also make use of the ``warped" metric $g_{mn} \equiv e^{-2A}
\tilde{g}_{mn}$.  The self-dual five-form flux $\tilde{F}_5 = *
\tilde{F}_5$ is
\begin{eqnarray}
\tilde{F}_5 = (1 + *)  \, d\alpha(y) \wedge \sqrt{-g_4} \, dx^0 \wedge dx^1 \wedge dx^2 \wedge dx^3 \,.
\end{eqnarray}
with $\alpha(y)$ a function on the compact space.
The NSNS and RR three-forms $H_3$, $F_3$ are combined with the complex RR axion/dilaton $\tau \equiv C_0 + i e^{-\phi}$ into the complex combination
\begin{eqnarray}
G_3 \equiv F_3 - \tau H_3 \,,
\end{eqnarray}
and we will find it useful to define the linear combinations
\begin{eqnarray}
\Phi_{\pm} \equiv e^{4A} \pm \alpha \,, \quad \quad G_\pm \equiv i G_3 \pm *_6 G_3\,,
\end{eqnarray}
where the six-dimensional Hodge star $*_6$  is the same using either $\tilde{g}_{mn}$ or $g_{mn}$.

In what follows we assume for simplicity a constant dilaton $\tau = i/g_s$.  The
Einstein equation and five-form
equation of motion can be
combined into the pair of equations
\begin{eqnarray}
- \tilde\nabla^2 (\Phi_\pm)^{-1} = {g_s^2 \over 96} \widetilde{|G_\pm|}^2 + 8 \pi^4 g_s \sum_{i_\pm} {\delta^6(y - y^{i_\pm}) \over \sqrt{\tilde{g}_6}} \,,
\label{PhiEqns}
\end{eqnarray}
where $i_\pm$ indexes the D3-branes/$\D3$-branes and the various tildes indicate that all contractions are with respect to the unwarped metric.  The
equations of motion for the three-forms are
\begin{eqnarray}
d (\Phi_+ G_-) = d (\Phi_- G_+) \,.
\end{eqnarray}

\subsection{Imaginary self-dual warped throat}

We will be concerned with backgrounds that are to leading order of the
Giddings-Kachru-Polchinski (GKP) \cite{GKP} type,
\begin{eqnarray}
\label{ISD} G_- = \Phi_- = 0 \, .
\end{eqnarray}
which in particular involve fluxes that are imaginary self-dual.\footnote{The relations (\ref{ISD}) will be
corrected by K\"ahler moduli stabilization effects and other
higher-order terms, and are explicitly violated by any
$\D3$-branes.}  This important class of solutions includes the
Klebanov-Strassler (KS) warped throat \cite{KS} arising as part of
a compact geometry, as described in \cite{GKP}.

To construct a KS throat, we begin with a conifold singularity
with corresponding three-cycles $A$ and $B$ and complex structure
modulus $\epsilon^2 = \int_A \Omega$,
with $\Omega$ the holomorphic
three-form.\footnote{Often one writes $\epsilon^2 = z$, but we
reserve the letter $z$ for something else.} The complex structure
modulus may then be stabilized by fluxes
\begin{eqnarray}
\int_A F_3 \equiv M \,, \quad \quad \int_B H_3 \equiv - K \,,
\end{eqnarray}
at an exponentially small value \cite{GKP}:
\begin{eqnarray}
\epsilon \sim e^{ - \pi K / g_s M} \,.
\end{eqnarray}
For $\epsilon \ll 1$, the resulting geometry is well-described by
a KS throat \cite{KS} over which the warp factor is strongly
varying, attached to the rest of the compact space where the warp
factor is mostly constant.  We will be working in the throat
region, whose classical solution is known precisely up to
corrections from the bulk of the geometry.  (In
\S\ref{BranePotSec} we will estimate the influence of these bulk
corrections, following \cite{Ofer}.)

The unwarped metric is that of the deformed conifold, which comes to
a smooth end at what we will call the ``bottom" or the ``tip" of the
throat.  Far from the bottom
but still in the throat,
the metric is approximately that of the ordinary conifold,
\begin{eqnarray}
\label{ConifoldMetric}
\tilde{g}_{mn} dy^m dy^n \approx dr^2 + r^2 ds^2_{T^{1,1}}  \quad{\rm for}  \quad r \gg \epsilon^{2/3} \,,
\end{eqnarray}
where $ds^2_{T^{1,1}}$ is the canonical metric on the
five-dimensional Einstein space $T^{1,1}$, which is topologically
$S^3 \times S^2$.  The warp factor in this region has the form
\begin{eqnarray}
\label{WarpFactor}
e^{-4A} = 2 (\Phi_+)^{-1} = {27 \pi \over 4 r^4} \left(g_s M
K\right) + \ldots \,,
\end{eqnarray}
where the dots denote logarithmic corrections, and we have set
$\alpha' = 1$.
At the tip of the throat the $S^2$ in $T^{1,1}$ shrinks to zero
size while the $S^3$ remains finite, and the metric becomes
\begin{eqnarray}
\tilde{g}_{mn} dy^m dy^n \approx \epsilon^{4/3} (d\tau^2 + \tau^2 d\Omega_2^2 + d \Omega_3^2) \,,
\end{eqnarray}
where we may parameterize the metric on the three-sphere as
\begin{eqnarray}
\label{3Sphere}
d\Omega^2_3 = (d\psi + \cos \theta d\phi)^2 + d\theta^2 + \sin^2 \theta d\phi^2 \,.
\end{eqnarray}
The radial coordinate $\tau$ is related to $r$ by
\begin{eqnarray}
r^3 = \epsilon^2 \cosh \tau \,,
\label{rtau}
\end{eqnarray}
and the tip is at $r^3 = \epsilon^2$, $\tau = 0$; the warp factor there is
\begin{eqnarray}
\label{a0def}
a_0 \equiv e^{A_0} \sim {\epsilon^{2/3} \over (g_s M)^{1/2}} \,,
\end{eqnarray}
up to numerical factors,
and is constant over the $S^3$.  Hence in the {\em warped} compact
metric the powers of the small number $\epsilon$ cancel, leaving
for the total metric near the bottom $\tau \approx 0$,
\begin{eqnarray}
\label{TipMetric}
ds^2 = {\epsilon^{4/3} \over g_s M} \, g_{\mu\nu} dx^\mu dx^\nu + (g_s M) (d\tau^2 + \tau^2 d\Omega_2^2 + d \Omega_3^2) \,,
\end{eqnarray}
where we have neglected the Weyl factor $e^{2u(x)}$ for
simplicity. Hence the three-sphere has radius
$R^2_{S^3} = g_s M$ in string units.
A necessary condition for the supergravity approximation to hold
is for this radius to be large, $g_s M \gg 1$.

\subsection{Four-dimensional action}
The effective theory for this warped compactification is a
four-dimensional ${\cal N}=1$ supergravity.  The scalar fields of
the effective theory fall into two categories: closed string
moduli, {\it{i.e.}}
the K\"ahler moduli, complex structure moduli, and dilaton; and
open string moduli, including the positions of any D-branes, which for
us will be D3-branes and D7-branes.

The flux superpotential,
\begin{eqnarray}
\label{GVWW}
W_0 \equiv \int G_3 \wedge \Omega \,,
\end{eqnarray}
gives rise to a potential for the dilaton and for the complex
structure moduli.
In addition, the F-theory generalization of (\ref{GVWW}), which includes 7-brane
worldvolume fluxes as well, can stabilize the positions of D7-branes.  We will make the assumption throughout that the flux-induced potential has stabilized all these moduli.

The remaining closed string moduli are then the K\"ahler moduli,
each of which is related to the volume of a four-cycle.  For
simplicity we will usually consider a single such modulus $\rho$,
whose real part is associated to the overall volume; in the
absence of brane fields we have
\begin{eqnarray} \label{equ:rhowithoutbranes}
\rho \equiv {1 \over 2} e^{4u} + i b \,,
\end{eqnarray}
where $b$ is the integral of the RR potential $C_4$ over the
corresponding four-cycle.

We will also be concerned with the dynamics of a mobile D3-brane
that fills spacetime and is pointlike in the compact space. We
denote the location of this D3-brane in the compact space by three
complex scalars $Y^I$, $I=1,2,3$.  In the presence of a D3-brane,
the real part of $\rho$ is related to the volume in a more
complicated way \cite{DeWolfeGiddings} than in (\ref{equ:rhowithoutbranes}),
\begin{eqnarray}
 \label{equ:rhowithbranes}
e^{4u} =  \rho + \bar\rho - k (Y,\bar{Y})/3 \,,
\end{eqnarray}
where $k$ is the geometric K\"ahler potential for the metric
on the Calabi-Yau: $\tilde{g}_{I \bar{J}} = \partial_I
\partial_{\bar{J}} k$.  The full K\"ahler potential for the D3-brane fields and the K\"ahler modulus $\rho$ is
\begin{eqnarray}
\label{BigKahler}
K = -3 \log e^{4u} = -3 \log (\rho + \bar\rho -  \gamma k (Y,\bar{Y})/3) \,,
\end{eqnarray}
with $\gamma = T_{D3} \kappa_4^2$.
For simplicity we will drop the constant $\gamma$, which is a pure
number when $\alpha'=1$, in the remainder of the paper; it may
always be restored by inserting it wherever $k(Y, \bar{Y})$
appears.

This K\"ahler potential generates kinetic terms matching those
derived from the Born-Infeld action for a D3-brane,
\begin{eqnarray}
\label{BI}
S_{D3} = - T_{D3} \int d^4x \sqrt{-g_4} e^{-4u}  g^{\mu\nu} \partial_\mu Y^I \partial_\nu \bar{Y}^{\bar{J}} \tilde{g}_{I \bar{J}} \,.
\end{eqnarray}
Notice that all powers of the warp factor have cancelled, but a
factor of $\epsilon^{4/3}$ is hiding in $\tilde{g}_{I \bar{J}}$.

We will also deal with $\D3$-branes; although these break
supersymmetry, their kinetic terms may still be written in the
form (\ref{BI}).  Both D3-branes and $\D3$-branes feel a potential
from a warped background with a nontrivial $\tilde{F_5}$; this is given by
\begin{eqnarray}
\label{BraneCoupling}
S_{{\rm D3}/\D3} = -T_{D3} \int d^4x \sqrt{-g_4} \, \Phi_\mp \,,
\end{eqnarray}
where the upper sign is for D3-branes and the lower for the
$\D3$-branes.  We see that D3-branes feel no force in our
leading-order imaginary self-dual background (\ref{ISD}).  Antibranes, in
contrast, feel a force from $\Phi_+$ (\ref{WarpFactor}) that draws them to the bottom
of the throat.

\subsection{Nonperturbative effects and K\"ahler moduli
stabilization}

Fluxes alone cannot stabilize the K\"ahler moduli in a type IIB
compactification.  Stable vacua will arise only in the presence of
some additional ingredient or mechanism.  The best-understood
mechanism of this sort is K\"ahler moduli stabilization due to
nonperturbative effects, as pioneered by
Kachru, Kallosh, Linde, and Trivedi (KKLT) \cite{KKLT}.

To each K\"ahler modulus $\rho_{i}$ is associated a holomorphic
four-cycle $\Sigma_4^{(i)}$.  The KKLT scenario requires that for
each K\"ahler modulus that is to be stabilized, a brane or branes
must be wrapped over the corresponding four-cycle. These wrapped
branes can be either a stack of $n>1$ D7-branes that also fill
four-dimensional spacetime, and on which strong gauge dynamics can
occur, or else a Euclidean D3-brane (D3-brane instanton).  Either
mechanism produces a $\rho_i$-dependent contribution to the
superpotential, generically leading to stabilization of the modulus
$\rho_i$.

For concreteness of presentation we refer to the
case with a single K\"ahler modulus $\rho$, and in which $n$ D7-branes wrap
$\Sigma_4$.  Our results extend trivially to the case of a
superpotential generated by Euclidean D3-branes,
for which one may put $n=1$.

The nonperturbative superpotential takes the form \beq W_{np} = A
\, e^{-a\rho} \,, \eeq where $a$ is a constant and the prefactor $A$,
which depends on the complex structure moduli, comes from
threshold corrections to the gauge coupling of the D7-brane. Ganor
\cite{Ganor} has given a topological argument that implies that in
fact $A$ also depends on the positions of any D3-branes in the
compactification.  This has been confirmed by explicit calculation
in toroidal orientifolds \cite{BHK} and in warped throat
backgrounds \cite{Baumann}; see also \cite{GiddMaha}.

Specifically, suppose that $y^1$, $y^2$, $y^3$ are three complex
coordinates in a region of the Calabi-Yau, and that $\Sigma_4$ is
defined by a single algebraic equation involving the $y^I$, which we
may write as $f(y)=0$. Recall that the $Y^I$ are also three complex
coordinates on the D3-brane configuration space, which is precisely
the Calabi-Yau manifold. The result of \cite{Baumann} is that the
prefactor $A$ is \beq A(Y) = A_0\, f(Y)^{1/n} \,, \eeq where $A_0$ depends
on the complex structure moduli but not on $Y$ or on $\rho$.  It
follows that the nonperturbative superpotential vanishes whenever
the D3-brane sits on the four-cycle $\Sigma_{4}$ wrapped by the
D7-branes.

We will find it convenient to define
\begin{eqnarray} \label{zetais}
\zeta(Y) \equiv -{1 \over n} \log f(Y) \,,
\end{eqnarray}
so that the total superpotential is \beq \label{equ:wwithzeta} W =
W_0 + A_0\, e^{-a\rho-\zeta(Y)} \,, \eeq where the constant $A_0$
captures the effects of the integrated-out complex structure
moduli.
In the remainder of the paper we will study the consequences of
the superpotential corrections $\zeta(Y)$.

\section{Supersymmetric vacua for D3-branes}
\label{SUSYVacSec}
\setcounter{equation}{0}

In this section, we present in full generality the conditions for
supersymmetric vacua of D3-branes in the presence of
nonperturbative moduli stabilization.  We reduce the equations for the locations of the
D3-branes to
those for the stationary points of a potential-type function
involving $k(Y,\bar{Y})$ and $\zeta(Y)$ only.  We also give the
equation for the stabilization of the K\"ahler modulus in terms
of the D3-brane location.

We turn first to the case of a compactification without D3-branes as a review, before confronting the case of interest.

\subsection{K\"ahler modulus alone}

Consider the nonperturbative superpotential for the single modulus
$\rho$,
\begin{eqnarray}
W = W_0 + A_0 \, e^{-a\rho} \,,
\end{eqnarray}
with $W_0$, $A_0$ complex constants and $a$ real.   The
F-term $DW \equiv \partial W + W \partial K$ is
\begin{eqnarray}
D_\rho W = - {e^{-a\rho} \over \rho + \bar\rho} [3 W_0 e^{a\rho} +
A_0 (3 + a \rho + a \bar\rho)] \,.
\end{eqnarray}
Supersymmetric solutions are found when this vanishes, leading to
\begin{eqnarray}
 e^{-a \rho} (3 + a \rho + a \bar\rho) = -{3 W_0 \over A_0} \,.
\end{eqnarray}
The overall volume $e^{4u}$ then solves the transcendental
equation
\begin{eqnarray}
e^{-a e^{4u}} (3 + a e^{4u})^2 = 9 \left| {W_0 \over A_0} \right|^2
\,, \label{RhoNoXEqn}
\end{eqnarray}
while the axion $b$ is fixed as
\begin{eqnarray}
b = -{1 \over a} \arg \left(- {W_0 \over A_0} \right) \,.
\label{NoBraneAxionFix}
\end{eqnarray}

\subsection{K\"ahler modulus and D3-brane}

Now add a D3-brane, which feels no potential in the absence of
K\"ahler moduli stabilization.  Its location is parameterized by
three complex coordinates $Y^I$.  The superpotential is now
(\ref{equ:wwithzeta}),
while the K\"ahler potential is as in (\ref{BigKahler}).
Recall first that the functional form of a K\"ahler potential is
always ambiguous up to the addition of a holomorphic function
$\xi$ and its conjugate $\bar{\xi}$.  This is reflected here by a corresponding
ambiguity in $\zeta(Y)$ and $\rho$, and an overall invariance of
$K$ and $W$ under the ``little K\"ahler
transformations'',
\begin{eqnarray} \nonumber
k &\to& k + 3~ \xi(Y) + 3~ \bar{\xi}(\bar{Y}) \,, \\
\label{LittleKahler}
\rho &\to& \rho + \xi(Y) \,, \\
\zeta &\to& \zeta - a \, \xi(Y) \,,
\nonumber
\end{eqnarray}
where it is clear that both $e^{4u}$ (\ref{equ:rhowithbranes}) and $a \rho + \zeta(Y)$ are invariants.

We now search for supersymmetric vacua.  The F-terms
that must vanish are
\begin{eqnarray} \label{equ:rhofterm}
D_\rho W &=& - a A_0 e^{-a\rho - \zeta(Y)} - { 3 W \over \rho + \bar\rho - k(Y,\bar{Y})/3} \,, \\
D_I W &=& -A_0\, \partial_I \zeta(Y) e^{-a \rho - \zeta(Y)} +
\partial_I k {W \over \rho + \bar\rho - k(Y,\bar{Y})/3} \,,
\end{eqnarray}
where we write $\partial_I \equiv \partial_{Y^I}$.  Eliminating $W/(\rho
+ \bar\rho - k/3)$ using the vanishing of $D_\rho W$, the vanishing
of $D_I W$ implies
\begin{eqnarray}
\label{YEqns}
\partial_I \zeta(Y) + {a \over 3} \partial_I k(Y,\bar{Y}) = 0 \,.
\end{eqnarray}
This is the sought-after equation for vacua for the
$Y^I$.  The vanishing of (\ref{equ:rhofterm}) then becomes
\begin{eqnarray}
\nonumber
- {3 W_0 \over A_0} &=& e^{-a \rho - \zeta(Y)}(3 + a \rho + a \bar\rho - a k(Y,\bar{Y})/3) \,, \\
&=& e^{-a \rho - \zeta(Y)} (3 + a e^{4u}) \,, \label{RhoEqn}
\end{eqnarray}
where we have used (\ref{equ:rhowithbranes}).
Multiplying (\ref{RhoEqn}) by its complex conjugate, we arrive at
an expression fixing $e^{4u}$ in terms of the already-determined
$Y^I$:
\begin{eqnarray}
e^{-a e^{4u}} (3 + a e^{4u})^2 &=& 9 \left| {W_0 \over A_0}
\right|^2 e^{\zeta(Y) + \bar\zeta(\bar{Y}) + a k(Y,\bar{Y})/3} \,.
\end{eqnarray}
This is manifestly little-K\"ahler-invariant, as the potential function
\begin{eqnarray}
{\cal V}(Y,\bar{Y}) \equiv \zeta(Y) + \bar\zeta(\bar{Y}) + a k(Y,\bar{Y})/3 \,,
\end{eqnarray}
is invariant as well; this is closely analogous to the standard ${\cal N}=1$
supersymmetry invariant ${\cal G} \equiv K + \log W + \log
\overline{W}$.
Note that we can combine the $\zeta$, $\bar{\zeta}$ with $A_0$ to
obtain
\begin{eqnarray}
\label{KahlerFix}
e^{-a e^{4u}} (3 + a e^{4u})^2 &=& 9 \left| {W_0 \over A(Y)}
\right|^2 e^{a k(Y,\bar{Y})/3} \,.
\end{eqnarray}
This equation is the exact analogue of (\ref{RhoNoXEqn}), the
equation for $e^{4u}$ in the case with no
D3-brane, with the addition
of the factor involving $k(Y,\bar{Y})$ to ensure little K\"ahler
invariance.  Similarly, the axion $b = {\rm Im}\ \rho$ is fixed as
\begin{eqnarray}
\label{AxionFix}
b = -{1 \over a} \arg \left(- {W_0 e^{\zeta(Y)} \over A_0} \right)
= -{1 \over a} \arg \left(- {W_0 \over A(Y)} \right) \,,
\end{eqnarray}
in precise analogy to (\ref{NoBraneAxionFix}).  Note also that we can write the desired equations  for the
D3-brane vacua (\ref{YEqns}) in a little-K\"ahler-invariant way:
\begin{eqnarray}
 d {\cal V}(Y, \bar{Y}) \equiv d \left[ \zeta(Y) + \bar\zeta(\bar{Y}) + a k(Y,\bar{Y})/3 \right]
 = 0 \,, \label{D3LocEqn}
\end{eqnarray}
where $d$ stands for either $\partial_{Y^I}$ or $\partial_{\bar{Y}^{\bar{I}}}$;
hence the D3-brane seeks to extremize the potential function
${\cal V}(Y, \bar{Y})$ that is the little-K\"ahler-invariant
generalization of the K\"ahler potential $k(Y, \bar{Y})$.

Equation (\ref{D3LocEqn}) is the general expression for the
D3-brane location in the presence of nonperturbative moduli
stabilization with a single K\"ahler modulus. It constitutes six real equations
in six real unknowns; consequently one would expect on general
grounds that generic solutions will entirely fix the D3-brane
moduli and localize the branes to points.

In the next section, we will consider a set of concrete examples
of vacua in the warped deformed conifold throat.  We shall find
that although some vacua are indeed points, it is also common
for examples that preserve some symmetry to leave one- or even
two-dimensional moduli spaces for the D3-branes.

\section{D3-brane potential at the tip of the deformed conifold}
\label{ConifoldVacSec}
\setcounter{equation}{0}

We will now study the D3-brane vacuum equation (\ref{D3LocEqn}) in
the warped deformed conifold, for various embeddings of the
wrapped D7-branes.
We will focus on the tip of the throat, although we briefly describe
off-tip results in Appendix C.

Several families of supersymmetric D7-brane embeddings are known in the (deformed) conifold throat.  In any given compact model, however, most of these embeddings are not realized: the global topology will select the possible
compact four-cycles, and then moduli stabilization will select fixed values for the location of the D7-brane on those cycles.  Here we will study two broad classes of embeddings, but we should bear in mind that any given compact model will strongly constrain the possibilities.

\subsection{Wrapped brane embeddings in homogeneous coordinates}

Each supersymmetric D7-brane embedding $f(Y)$ is defined in terms
of one of the two natural sets of homogeneous variables $z^A$, $w_i,  ~ A, i
= 1 \ldots 4$ that define the deformed conifold:
\begin{eqnarray}
\label{ConifoldEqn} \sum_{A=1}^4 (z^A)^2 =-2 (w_1w_2 - w_3 w_4) =
\epsilon^2 \,,
\end{eqnarray}
either as $f(z^A)$ or $f(w_i)$.  It will be convenient for us to evaluate the D3-brane vacuum equation (\ref{D3LocEqn}) in terms of whichever variables appear in $f$.
Later in the section we shall pass to angular coordinates on the $S^3$ at the tip, which are more intuitive for describing that locus.

These homogeneous variables transform under the $SO(4) \sim
SU(2)_L \times SU(2)_R$ isometry that acts on the conifold: the
$z^A$ are a ${\bf 4}$ of $SO(4)$ and the $w_i$ are a $({\bf 2},
{\bf 2})$ of $SU(2)_L \times SU(2)_R$.  Many
D7-brane embeddings
preserve a subgroup of $SO(4)$, requiring the D3-brane moduli
space to fill out an orbit of this preserved symmetry.  Thus in cases with symmetry-preserving D7-branes, D3-brane vacua will either sit at a fixed point of the preserved isometries or
else occupy a continuous moduli space; we will find that both cases are indeed realized.

In making use of the homogeneous variables, we must take into account that they are an overcomplete set satisfying the constraint (\ref{ConifoldEqn}).  A straightforward way to do this is to
eliminate one variable in terms of the others, which is akin to choosing a gauge.  Doing this for any given variable is in general not valid for the entire coordinate range, so we will need to consider different gauge choices to see all vacua.  For example, we may eliminate the first variable in each case:
\begin{eqnarray}
\label{VarElim}
z^1 = \sqrt{\epsilon^2 - (z^2)^2 - (z^3)^2 - (z^4)^2}  \,, \quad \quad w_1 = {w_3 w_4 - \epsilon^2/2 \over w_2}\,.
\end{eqnarray}
The derivatives in terms of the independent variables $z^2, z^3, z^4$ and $w_2, w_3, w_4$ are then
\begin{eqnarray}
\label{zDeriv}
{\partial z^1 \over \partial z^a} = - z^a (\epsilon^2 - (z^2)^2 - (z^3)^2 - (z^4)^2)^{-1/2}  = - {z^a \over z^1} \,,
\end{eqnarray}
where $a=2,3,4$, and
\begin{eqnarray}
{\partial w_1 \over \partial w_2} = -  {(w_3 w_4 - \epsilon^2/2) \over w_2^2} = - {w_1 \over w_2} \,, \quad\quad
{\partial w_1 \over \partial w_3} = {w_4 \over w_2} \,, \quad \quad
{\partial w_1 \over \partial w_4} = {w_3 \over w_2} \,.
\end{eqnarray}
We note that eliminating the $z^1$ variable forces us to use a patch where $z^1 \neq 0$, while eliminating $w_1$ forces us to a patch with $w_2 \neq 0$.

\subsection{Vacua at the tip}
\label{VacuaTipSec}

We shall concern ourselves primarily with D3-brane vacua at the tip of the conifold.  At this locus the homogeneous variables are further constrained by
\begin{eqnarray}
\label{rhomog}
r^3 \equiv \sum_{A=1}^4 |z^A|^2 \equiv \sum_{i=1}^4 |w_i|^2 = \epsilon^2 \,,
\end{eqnarray}
which combined with (\ref{ConifoldEqn}) implies
the relations
\begin{eqnarray}
\label{TipConstraints}
w_1 = - \overline{w}_2 \,, \quad w_3 = \overline{w}_4 \,, \quad \quad z^A = \bar{z}^A \,, \quad \quad {\rm at\ the\ tip}\,,
\end{eqnarray}
as is detailed in Appendix A.  Note that since $|w_1| = |w_2|$ on the tip, our patch with $w_2 \neq 0$ also has $w_1 \neq 0$ there.

In principle, solutions to (\ref{D3LocEqn}) may involve
non-constant $\zeta(Y)$ playing off against non-constant $k(Y,
\bar{Y})$.  However, the geometric K\"ahler
potential depends only on the radial variable $\tau$ and
has the form near the bottom of the throat \cite{Candelas},
\begin{eqnarray}
\label{KahlerPot}
\partial_\tau k = 2^{-1/2} \epsilon^{4/3} (\sinh 2\tau - 2 \tau)^{1/3} \sim \frac{2^{1/6}}{3^{1/3}} \, \epsilon^{4/3} \, \tau + {\cal O}(\tau^3) \,,
\end{eqnarray}
which vanishes at the tip; hence the K\"ahler potential $k$ is stationary in all directions there. Equivalently, we may show that $\partial_{z^A} k = \partial_{w_i} k = 0$ at the tip.  Using (\ref{rtau}) and (\ref{rhomog}), we get from (\ref{KahlerPot}),
\begin{eqnarray}
k = k_0 +  \frac{2^{1/6}}{3^{1/3}} \, \epsilon^{-2/3} \left( \sum_{A=1}^4 |z^A|^2 - \epsilon^2 \right) \,,
\end{eqnarray}
with $k_0$ a constant.  The derivative with respect to $z^a$, $a = 2,3,4$ is then
\begin{eqnarray}
\partial_{z^a} k   \propto \left(\frac{\partial z^1}{\partial z^a} \bar{z}^1 + \bar{z}^a \right) = \left( - {z^a \bar{z}^1 \over z^1} + \bar{z}^a \right) \,,
\end{eqnarray}
which vanishes at the tip due to the $z^A$ being real there, (\ref{TipConstraints}).  An analogous computation (or the chain rule) shows $\partial_{w_i} k = 0$ at the tip as well.

Thus we see
that the term involving the K\"ahler potential drops out of (\ref{D3LocEqn})
at the tip of the throat.
Consequently, finding D3-brane vacua at that locus then reduces simply to solving
\begin{eqnarray}
\label{ThroatBraneEqn}
\partial_{Y^I} \zeta = 0 \,,
\end{eqnarray}
or, using (\ref{zetais}),
\begin{eqnarray}
\label{D3Eqn2}
\partial_{Y^I} \log f(Y) = {\partial_{Y^I} f(Y)\over f(Y)} =0 \,.
\end{eqnarray}
In the remainder of the section we will focus on solutions to (\ref{D3Eqn2}).

We notice immediately that when the D3-brane sits on the
D7-brane, the denominator in (\ref{D3Eqn2}) vanishes, preventing
(\ref{D3Eqn2}) from being satisfied unless the numerator were to vanish even more rapidly.
In the examples to follow
the numerator will not vanish quickly enough, and we will  conclude that in general, the D3-branes will be confined to a locus {\em away} from the moduli-stabilizing wrapped D7-branes.
Additionally, we see that the D3-brane vacua must be
`symmetrically oriented' with respect to the wrapped branes, in
the sense that the D3-branes sit at an extremum of $f$.

\subsection{D3-brane potential of ACR embeddings}
\label{ACRSec}

\subsubsection*{General solutions for ACR embeddings}

An infinite class of holomorphic four-cycles that admit
supersymmetric wrapped D7-branes was found by Are\'an, Crooks, and
Ramallo (ACR) \cite{ACR},

\begin{eqnarray} \label{ACRembedding}
f(w_i) = \prod_{i=1}^4 w_i^{p_i} - \mu^P \,,
\end{eqnarray}
where $p_i$ are four integers and $P \equiv\sum_{i=1}^4 p_i$.  The D3-brane vacua
(\ref{D3Eqn2}) will occur  where the derivatives,
\begin{eqnarray}
\label{w2deriv}
{\partial f \over \partial w_2} &=& = (p_2 - p_1) w_1^{p_1} w_2^{p_2-1} w_3^{p_3} w_4^{p_4} \,, \\
{\partial f \over \partial w_3} &=& = (p_1 w_3 w_4 + p_3 w_1 w_2) w_1^{p_1-1} w_2^{p_2-1} w_3^{p_3-1} w_4^{p_4} \,,\label{w3deriv} \\
{\partial f \over \partial w_4} &=& = (p_1 w_3 w_4 + p_4 w_1 w_2) w_1^{p_1-1} w_2^{p_2-1} w_3^{p_3} w_4^{p_4-1} \,,
\label{w4deriv}
\end{eqnarray}
are set to zero,
and also $f(w_i) \neq 0$.

We now find the general solution for D3-brane vacua at the bottom of the throat in this class of D7-brane embeddings.  Recall that in our coordinate choice, we must have $|w_1| = |w_2| \neq 0$ at the tip.  Setting the derivative (\ref{w2deriv}) to vanish thus requires either:
\begin{eqnarray}
p_1 = p_2 \,, \quad \quad {\rm or}\ \quad \quad w_3 = w_4 = 0\,, \;\; p_3 + p_4 \geq 1\,.
\end{eqnarray}
Meanwhile the second two derivatives (\ref{w3deriv}), (\ref{w4deriv}) can be set to zero either by
\begin{eqnarray}
w_3 = w_4 = 0 \,, \;\; p_3 + p_4 \geq 2 \,,
\end{eqnarray}
or
\begin{eqnarray}
p_3 = p_4 \,, \quad p_1 w_3 w_4 + p_3 w_1 w_2 \equiv p_1 |w_3|^2 - p_3 |w_1|^2 = 0\,.
\end{eqnarray}
The possible solutions are as follows.  First, clearly $w_3 = w_4 = 0$ is a solution of all equations for $p_3 + p_4 \geq 2$. If $p_3 + p_4 = 1$ we may set (\ref{w2deriv}) to zero with $w_3 = w_4 = 0$, but then the other equations have no solution, so this must be discarded.  All other solutions thus have $p_1 = p_2$ to set (\ref{w2deriv}) to zero,
and $p_3 = p_4$ and $p_1 |w_3|^2 = p_3 |w_1|^2$ so that (\ref{w3deriv}) and (\ref{w4deriv}) vanish.

This is the complete set of ACR vacua in the gauge where $w_1$ and $w_2$ must not vanish.  However, there are also vacua with $w_1 = w_2 = 0$ that may be found by repeating the same analysis in the gauge where $w_3$ is eliminated as an independent variable, and  $1 \leftrightarrow 3$, $2 \leftrightarrow 4$ in the equations above.

We may thus summarize the three classes of D3-brane vacua in ACR embeddings as,
\begin{eqnarray}
{\rm A}. && w_1 = w_2 = 0 \quad {\rm if} \ \quad p_1 + p_2 \geq 2\,,\\
{\rm B}. && w_3 = w_4 = 0 \quad {\rm if} \ \quad p_3 + p_4 \geq 2\,, \\
{\rm C}. && p |w_3|^2 = q |w_1|^2 \quad {\rm if} \ \quad p_1 = p_2 \equiv p \,, \;\; p_3 = p_4 \equiv q \,.\nonumber
\end{eqnarray}
To understand the dimensionality of these loci, it is convenient to translate them into coordinates $\psi, \theta, \phi$ on the $S^3$ (\ref{3Sphere}).  As derived in Appendix A, $w_1 = w_2 = 0$ corresponds to $\theta = 0$, while $w_3 = w_4 = 0$ is $\theta = \pi$.  It is useful to define angles $\alpha \equiv \psi + \phi$, $\beta \equiv \psi - \phi$, in terms of which the metric on the
three-sphere (\ref{3Sphere}) becomes
\begin{eqnarray}
\label{MetricAB} ds^2 = d\theta^2 + \cos^2\left( {\theta \over
2}\right) d\alpha^2 + \sin^2\left( {\theta \over 2}\right)
d\beta^2 \,.
\end{eqnarray}
Thus, at each of $\theta = 0$ and $\theta = \pi$, one of the
angles $\alpha, \beta$ becomes degenerate and the other parameterizes
an $S^1$.  Hence solutions of classes A and B are both $S^1$'s.

Meanwhile, class C vacua occur for $p_1 = p_2 \equiv p$ and $p_3 = p_4 \equiv q$.    Obviously solutions to $p |w_3|^2 = q |w_1|^2$ only exist when $p$ and $q$ do not have opposite signs; in terms of the $S^3$, this is solved by $\theta  = 2 \tan^{-1} \sqrt{p /q}$.  If either $p=0$ or $q=0$, this solution reduces to $\theta = 0$ or $\theta  = \pi$ and is an $S^1$, just like solutions A or B.  For both $p$ and $q$ nonvanishing and of the same sign, neither $\alpha$ nor $\beta$ is degenerate, and the locus of vacua is two-dimensional, with (\ref{MetricAB}) indicating the space is a $T^2$.

The vacua of the ACR embeddings are thus all in one- or two-dimensional, continuous moduli spaces:
\begin{eqnarray}
\label{FirstSoln}
{\rm A}. && S^1: \quad \theta = 0 \quad {\rm if} \ \quad p_1 + p_2 \geq 2\,,\\
{\rm B}. && S^1: \quad \theta = \pi \quad {\rm if} \ \quad p_3 + p_4 \geq 2\,,\\
{\rm C1}. && T^2: \quad \theta = 2 \tan^{-1} \sqrt{p/q} \quad {\rm if} \ \quad p_1 = p_2 \equiv p \neq 0 \,,\;\; p_3 = p_4 \equiv q \neq 0\,, \\
{\rm C2}. && S^1: \quad \theta = 0 \quad {\rm if} \ \quad p_1 = p_2 = 0 \,,\;\; p_3 = p_4 \neq 0\,, \\
{\rm C3}. && S^1: \quad \theta = \pi \quad {\rm if} \ \quad p_1 = p_2 \neq 0 \,,\;\; p_3 = p_4 = 0\,.
\label{LastSoln}
\end{eqnarray}
These solutions may fail to exist if $f(z^a) = 0$ happens to
coincide with the putative vacuum locus.  This in general depends on the value of $\mu$ in the embedding (\ref{ACRembedding}).  For solutions of type A or B, this
occurs for $\mu=0$.  For type C solutions, it occurs when
\begin{eqnarray}
\left({\sqrt{2} \mu \over \epsilon}\right)^P = \left( - p \over p+q\right)^p \,\left( q \over p+q\right)^q \,,
\end{eqnarray}
where we used that in this case $P=2p+2q$.  In particular, $\mu^P$ must be real for this to hold.  For generic values of $\mu$, the D7-brane will not intersect the moduli spaces and the solutions will exist.

\subsubsection*{The Karch-Katz embedding}

As our first example, consider the {\em Karch-Katz embedding}  \cite{KarchKatz}.  It has the form
\begin{eqnarray}
f_{KK} = w_1 w_2 - \mu^2 \,.
\label{eq:KarchKatzEmbed}
\end{eqnarray}
Hence $p_1 = p_2 = 1$ and $p_3 = p_4  =0$.  We readily see that it has two types of solutions: one of type A (\ref{FirstSoln}), an $S^1$ at $\theta = 0$, and one of type C3 (\ref{LastSoln}), also an $S^1$, this one at $\theta = \pi$.

This solution preserves an $SO(2) \times SO(2)$ subgroup of $SO(4)$.  These $SO(2)$'s may be thought of as the phases in $w_1/w_2$ and $w_3/w_4$.  At the tip, they are realized precisely as shifts of the angles $\alpha \equiv \psi + \phi$ and $\beta \equiv \psi - \phi$, respectively.  At each locus, $\theta = 0$ or $\theta = \pi$, one of the two angles is degenerate while the other parameterizes the $S^1$; hence one of the $SO(2)$'s is faithfully represented, while the other is trivial, on each component of the moduli space.

It is interesting to note that the Karch-Katz embedding may also be written as
\begin{eqnarray}
f_{KK} = - {(z^1)^2 + (z^2)^2 \over 2} - \mu^2 \,.
\end{eqnarray}
The $SO(2)$'s are then rotations of $z^1$ into $z^2$, and
of $z^3$ into $z^4$.

\subsubsection*{Generalized Karch-Katz embeddings}

It is natural to categorize D7-brane embeddings by their preserved symmetry.  The ones that preserve $SO(2) \times SO(2)$, as
the Karch-Katz embedding does, are precisely those that are functions of
$w_1 w_2 = -  \left((z^1)^2 + (z^2)^2\right)/2$ and $w_3 w_4 =   \left((z^3)^2 + (z^4)^2\right)/2$; this is the condition $p_1= p_2 \equiv p$, $p_3 = p_4 \equiv q$, or precisely the circumstances where solution C may appear (solution C also requires ${\rm sgn}(p) = {\rm sgn}(q)$).  This is thus the class of ACR embeddings that may have two-dimensional moduli spaces.

As an example, consider
\begin{eqnarray}
f = w_1 w_2 w_3 w_4 - \mu^4 \,,
\end{eqnarray}
with $p_1 = p_2 = p_3 =p_4 = 1$.  This embedding has an $S^1$ moduli space of type A, an $S^1$ moduli space of type B, and a $T^2$ moduli space of type C1 at $\theta = \pi/2$.  This last space is a square torus with $SO(2) \times SO(2)$ acting naturally on the two directions.

\subsubsection*{The Ouyang embedding}

As a final example, let us consider the {\em Ouyang
embedding} \cite{Ouyang}, which has the form $p_1 = 1$ with
the other $p_i$ vanishing; that is, the embedding function $f$ has the form
\begin{eqnarray}
f_O = w_1 - \mu \,.
\end{eqnarray}
This embedding breaks $SO(4) \to SO(2)$, the preserved symmetry being the relative phase of $w_3$ and $w_4$, which is $\alpha = \psi + \phi$ on the $S^3$ at the tip.

One can quickly see that {\em none} of the conditions enumerated in (\ref{FirstSoln})-(\ref{LastSoln}) hold in this case.  Consequently, the Ouyang embedding has {\em no} supersymmetric D3-brane vacua on the tip of the throat.  We see that it is not the case that all ACR embeddings have supersymmetric vacua there.

One naturally wonders whether there are supersymmetric vacua elsewhere in the throat, found by canceling the non-vanishing of the derivative of the embedding function against the non-vanishing of the K\"ahler potential.  In Appendix C we show that
generically there will indeed be such vacua; although we do not examine the issue, one may surmise that other embeddings will also have off-tip vacua.

\bigskip

Thus far we have found embeddings with moduli spaces of D3-brane vacua at the tip, as well as embeddings with no D3-brane vacua there.  However, so far we have not found an example of what we might think would be the most generic case --
a vacuum in which the D3-brane is stabilized at a point.  We now turn to a class that does possess such vacua, where the D3-brane position is completely stabilzed.

\subsection{D3-brane potential of Kuperstein embeddings}

\subsubsection*{General solutions for Kuperstein embeddings}

Not all supersymmetric embeddings of D7-branes fall into the class analyzed in the previous subsection.   Another set was described by Kuperstein \cite{Kuperstein}, having the form
(up to $SO(4)$ permutations of the $z^A$),
\begin{eqnarray}
\label{GenKuper}
f = z^1 - g\left((z^3)^2 + (z^4)^2\right) \,,
\end{eqnarray}
where $g(x)$ is a holomorphic function of $x \equiv (z^3)^2 + (z^4)^2$.  This class generally breaks $SO(4) \to SO(2)$, where the remaining symmetry is rotations preserving $x$.

We may classify general solutions of this class.  The derivatives of $f$ that must vanish are
\begin{eqnarray}
{\partial f \over \partial z^2} &=& - {z^2 \over z^1} \,, \\
{\partial f \over \partial z^3} &=& - {z^3 \over z^1} \left(1 + 2 z_1 g'(x)\right)\,, \\
{\partial f \over \partial z^4} &=& - {z^4 \over z^1}\left(1 + 2 z_1 g'(x)\right) \,.
\end{eqnarray}
One solution is obviously $z^2 = z^3 = z^4 = 0$; this holds so long as $g'(x=0)$
does not diverge.  The other is $z^2 = 0$, $\left(1 + 2 z_1 g'(x)\right) = 0$.  At $z^2 = 0$ we have $z^1= \sqrt{\epsilon^2 - x}$, so the solutions are
\begin{eqnarray}
\label{DSoln}
{\rm D}. && z^2 = z^3 = z^4 = 0  \quad {\rm if} \ \quad g'(x) \ {\rm finite} \,, \\
{\rm E}. && z^2 = 0\,, \ 1 + 2 \sqrt{\epsilon^2 - x} \, g'(x) = 0 \,.
\end{eqnarray}
In principle, one should also analyze (\ref{GenKuper}) in gauges where $z^2$ or one of $z^3$, $z^4$ is eliminated.  However, the former yields no solutions, while the latter again produces solution E.

Again, to discuss the solutions in more detail, we translate them into angular coordinates on the $S^3$ at the tip.  Solution D is a single point with $z^1 = \epsilon$; it translates into $\theta = \beta = \pi$.  (Note that $z^1 = -\epsilon$ is not independent since the defining relation of the conifold identifies the two points related by a flip in the sign of all four $z^A$.)

Meanwhile, for solution E a constraint is placed on $x = \epsilon^2 \cos^2 (\theta/2)$, while $z^2 = 0$ implies $\beta = \pi$.  The resulting space is an $S^1$ parameterized by $\alpha$; the $SO(2)$ symmetry of the embeddings corresponds to $\alpha$-rotations. If $x=0$, we have $\theta = \pi$ and the solution degenerates to the point solution D.

Thus we summarize,
\begin{eqnarray}
{\rm D}. &&{\rm Point}: \quad  \beta= \theta = \pi  \quad {\rm if} \ \quad g'(x=0) \ {\rm finite} \,, \\
{\rm E}. && S^1: \quad \beta = \pi\,, \ \theta = 2 \cos^{-1} (x/\epsilon^2)  \quad {\rm if} \ \quad x \neq 0  \; \;{\rm solves} \ \; 1 + 2 \sqrt{\epsilon^2 - x} \, g'(x) = 0 \,.
\end{eqnarray}
As with the ACR cases, vacua may fail to exist when $f=0$
happens to coincide with the candidate vacuum locus; this always depends on the constant term in the embedding function.  Solution D does not hold if $g(0) = \epsilon$, while solution E fails
if $2 g(x) g'(x) = -1$ for the stabilized value of $x$.

We see that completely stabilized D3-brane moduli are possible in this embedding, along with continuous moduli spaces.  We now turn to examples.

\subsubsection*{Simplest Kuperstein embedding}

The simplest case, which was focused on in \cite{Kuperstein} and
could be called simply the {\em Kuperstein embedding}, is defined by $g = const$,
or in other words
\begin{eqnarray}
\label{KupersteinEmbed}
f_K = z^1 - \mu \,,
\end{eqnarray}
for some complex $\mu$.  This embedding has a larger preserved symmetry than the general class, breaking $SO(4) \to SO(3)$, where $SO(3)$ acts naturally on the other three $z^a$.

Examining the possible vacua, we find that there is indeed a solution of type D (\ref{DSoln}), but since $g'(x) = 0$, no solution of type E.  Thus the only supersymmetric vacuum on the tip for this embedding is a single point with all moduli stabilized.

\subsubsection*{Kuperstein embedding with moduli space}

An elementary example in the Kuperstein class (\ref{GenKuper}) that has an $S^1$ moduli space is given by the linear function $g(x) = - x /(2 s \epsilon) +\mu$ where $s > 0$ is a constant; we have inserted the $2 \epsilon$ for convenience.  Then
\begin{eqnarray}
f = z^1 + {(z^3)^2 + (z^4)^2 \over 2 s \epsilon} - \mu \,.
\end{eqnarray}
Since $g'(x=0) = -1/(2 s \epsilon)$ does not diverge, the pointlike solution of type D exists at $z^2 = z^3 = z^4 = 0$.  There is also a solution of type E, having $\beta = \pi$ and satisfying $1 + 2 \sqrt{\epsilon^2-x} g'(x) = 1-  \sqrt{\epsilon^2 - x}/(s \epsilon) = 0$, solved by $x = \epsilon^2 (1-s^2)$, or $\sin (\theta/2) = s$.
The solution of type E only exists for $0 < s \leq 1$; it is generically an $S^1$ but coincides with the pointlike type D solution for $s=1$.

\bigskip

To summarize, we have found that the ACR and Kuperstein classes
of embeddings have a variety of possible D3-brane vacua, encompassing two- and one-dimensional moduli spaces such as $T^2$ and $S^1$, as well as discrete points.
There is no guarantee that a fixed compact model will realize any particular embedding; additionally, as we discuss in
\S\ref{InflationSec}, the bulk of a compact geometry will generally break the preserved isometries and lift the moduli spaces, though if the throat is strongly warped this breaking will be suppressed.  It is also possible that other supersymmetric embeddings may exist; however, in studying these cases we have uncovered a broad spectrum of possible vacua, and it is
plausible that any other embeddings will possess similar characteristics to the ones studied here.

\section{Vacua for $\D3$-branes}
\label{AntibraneSec}
\setcounter{equation}{0}

The $\D3$-brane feels a force from the flux background even in the
absence of the K\"ahler modulus stabilization.  It couples to
$\Phi_+$ as given in equation (\ref{BraneCoupling}), and the
background variation (\ref{WarpFactor}) of $\Phi_+$ draws the
$\D3$-brane to the bottom of the throat.  Because the warp factor
is independent of the coordinates on the $S^3$ at the bottom of
the throat, the leading-order background does not prefer one point
on the $S^3$ over another.

In addition, however, a $\D3$-brane feels a force from moduli
stabilization, as a D3-brane does. In this
section, we will explore the possible vacua for $\D3$-branes at
the tip. These vacua are all non-supersymmetric, of course, and at
best metastable \cite{FluxAnnihilation}.

The effect of moduli-stabilization forces on D3-brane motion
during slow-roll D-brane inflation is reasonably well-understood
\cite{KKLMMT}. However, the role of moduli-stabilization forces in
the final brane-antibrane annihilation stage has received
relatively little attention.  Since annihilation of a
brane-antibrane pair is the typical exit from inflation in this
class of models, it is important to ascertain whether the moduli
stabilization forces even
permit branes and antibranes to reach the same loci, or whether
they create an additional potential barrier to annihilation.  We
will find that at the tip of the throat, the two kinds of branes
do indeed have vacua at the same locations, so there is no
problematic barrier that could forestall tachyon condensation.

In the D3-brane case, the moduli stabilization force was derived
\cite{Baumann} by considering the backreaction of the brane on the
warp factor and on $C_4$; these determined the real and imaginary
parts of $\zeta(X)$, respectively.  The $\D3$-brane has by
definition the same source for gravity but the opposite $C_4$
charge; therefore going from brane to antibrane is realized by
flipping the sign of the imaginary part of $\zeta$, or in other
words, exchanging $\zeta(Y) \leftrightarrow \bar\zeta(\bar{Y})$ in
the potential:
\begin{eqnarray}
\label{AntiPot}
V_{\D3}(\rho, \bar\rho; Y, \bar{Y}) = V_{{\rm D3}} |_{\zeta(Y) \leftrightarrow \bar\zeta(\bar{Y})} \,.
\end{eqnarray}
The D3-brane potential, being supersymmetric, possesses minima
whose locations are determined by the first-order equations $DW =
0$.  A nonsupersymmetric potential like $V_{\D3}$, on the other
hand, in general has minima determined simply by $dV=0$, which in
terms of quantities like $\zeta$ and $k$ will generally be
second-order equations.  We may nonetheless ask: is it possible
that the $\D3$-brane vacua solve first-order equations analogous
to those of the D3-brane case?

Because of its relationship (\ref{AntiPot}) to the D3-brane
potential, we may write $V_{\D3}$ as
\begin{eqnarray}
V_{\D3} = e^K \left( (K^{-1})^{\alpha \bar{\beta}} \hat{F}_\alpha \hat{\overline{F}}_{\bar{\beta}} - 3 \hat{W} \hat{\overline{W}} \right) \,,
\end{eqnarray}
with
\begin{eqnarray}
\hat{F}_\alpha \equiv D_\alpha W |_{\zeta(Y) \leftrightarrow \bar\zeta(\bar{Y})}\,, \quad \quad \hat{W} \equiv W |_{\zeta(Y) \leftrightarrow \bar\zeta(\bar{Y})} \,,
\end{eqnarray}
and the indices $\alpha, \beta$ running over $\rho$ as well as $I = 1,2,3$ for the D3-brane.
This form obviously resembles a supersymmetric potential. The vanishing of the $\hat{F}_\alpha$, however, in general does not produce an extremum of the potential.  In calculating $d_\alpha V$, derivatives acting on $e^K$ and $\hat{W}$ together lead to terms in $D_\alpha \hat{W} \equiv \partial_\alpha \hat{W} + W \partial_\alpha K$; while this coincides with $\hat{F}_\alpha$ for $\alpha=\rho$, it does not in general for $\alpha = I$:
\begin{eqnarray}
\hat{F}_\rho = \partial_\rho \hat{W} + \hat{W} \partial_\rho K = D_\rho \hat{W} \,, \quad \quad
\hat{F}_I  = \partial_{\bar{Y}^{\bar{I}}} \hat{W} + \hat{W} \partial_{Y^I} K \neq D_I \hat{W} \,,
\label{FHat}
\end{eqnarray}
because of the exchange of $\zeta(Y)$ and $\bar\zeta(\bar{Y})$.
This is not surprising, as a non-supersymmetric potential has in
principle no reason to satisfy
equations of the sort $DW=0$.

However,
there is a special case where $\hat{F}_I$ and $D_I
\hat{W}$ will vanish simultaneously: when $\partial_{Y^I} K = 0$.  In
that case, $D_I \hat{W} = \hat{F}_{\bar{I}}$ and the vanishing of
the $\hat{F}_\alpha$ also implies the vanishing of the $D_\alpha \hat{W}$.
In this case, first-order equations do lead to a minimum of
$V_{\D3}$.

While this additional condition is not universal, as shown in
\S\ref{VacuaTipSec} on the tip of the $S^3$ we do indeed have
\begin{eqnarray}
\label{DKZero}
\partial_{Y^I} K \propto \partial_{Y^I} k  = 0 \,.
\end{eqnarray}
Consequently,
at the tip of the throat (or in any locus in a more complicated geometry where (\ref{DKZero}) holds), the $\D3$-brane potential
has a minimum for
\begin{eqnarray}
\hat{F}_\alpha \equiv D_\alpha W |_{\zeta(Y) \leftrightarrow \bar\zeta(\bar{Y})} = 0 \,.
\end{eqnarray}
Hence we may take the first-order equations for the D3-brane minima from the previous sections, and simply switch $\zeta(Y)$ with its conjugate to find the equations for minima for the $\D3$-branes.  Most important for us is that the equation (\ref{D3LocEqn}), which becomes
\begin{eqnarray}
d \left[ \zeta(Y) + \bar\zeta(\bar{Y}) \right]
= 0 \,,
\end{eqnarray}
is invariant under this conjugation.  Consequently, {\em the
$\overline{D3}$-branes are localized to the same subspace on the
$S^3$ as the D3-branes.}

We see additionally that the equation (\ref{KahlerFix}) for the
volume is unchanged.  The axion stabilization equation
(\ref{AxionFix}) is modified in a straightforward way,
fixing
the axion at a different value; because of the axion shift
symmetry that was present before the moduli stabilization
potential appeared, however, all frozen values of the axion are
physically equivalent.

Another way to see the same result is to consider the explicit form of the D3-brane potential and examine the transformation of each term under $\zeta(Y) \leftrightarrow \bar\zeta(\bar{Y})$.  The scalar potential for a brane can be written,
\begin{eqnarray}
V_{\rm D3} &=& e^{K}\left((K^{-1})^{\alpha \bar\beta}F_{\alpha}\overline{F}_{\bar\beta} - 3 |W|^2\right) \,, \\
& = & e^K \left( (K^{-1})^{\rho\bar\rho}F_{\rho}\overline{F}_{\bar{\rho}} + (K^{-1})^{\rho\bar{J}}F_{\rho}\overline{F}_{\bar{J}} +(K^{-1})^{I \bar\rho}F_{I}\overline{F}_{\bar{\rho}} + (K^{-1})^{I \bar{J}}F_{I}\overline{F}_{\bar{J}} - 3 |W|^2 \right) \,. \nonumber
\end{eqnarray}
We recall that the no-scale structure of the background implies
\begin{eqnarray}
(K^{-1})^{\alpha \bar{\beta}} \partial_\alpha K \partial_{\bar{\beta}} K - 3 = 0 \,,
\end{eqnarray}
for $\alpha = \rho, I$, causing all $|W|^2$ terms to vanish.  In addition, explicit calculation reveals that $(K^{-1})^{\rho \bar{J}}$ and $(K^{-1})^{I \bar{\rho}}$ are proportional to $\partial_{Y^I} K$ or $\partial_{\bar{Y}^{\bar{J}}} K$.  Hence the potential becomes
{\small
\begin{eqnarray}
V_{\rm D3}= e^K \left( (K^{-1})^{\rho\bar\rho} (\partial_\rho W \partial_{\bar\rho} \overline{W} + 2 {\rm Re}(\overline{W} \partial_{\bar\rho} K \partial_\rho W) ) + (K^{-1})^{I \bar{J}} \partial_I W \partial_{\bar{J}} \bar{W} + \partial_{Y^I} K (\ldots) +  \partial_{\bar{Y}^{\bar{J}}} K (\ldots)\right) \,
\end{eqnarray}}
The only one of the first three terms not invariant under the conjugation of $\zeta(Y)$ is the second.  The non-invariant term also happens to be the only one containing the axion ${\rm Im}\ \rho$:
\begin{eqnarray}
\label{NonInvar}
2 \, e^K (K^{-1})^{\rho\bar\rho}\,  {\rm Re}(\overline{W} \partial_{\bar\rho} K \partial_\rho W)  = - 2a \,e^K \, (K^{-1})^{\rho\bar\rho}\, {\rm Re}  (\overline{W}_0 A \, \partial_{\bar\rho}K e^{-a\rho - \zeta(Y)}) + \ldots \,.
\end{eqnarray}
We find that in transforming to $V_{\D3}$ from $V_{\rm D3}$, provided $\partial_{Y^I} K = 0$ as is the case at the throat's tip, the only change in the potential to consider is from the term (\ref{NonInvar}).  However, we observe that this change -- the flip in the sign of ${\rm Im}\ \zeta(Y)$ -- can be compensated for simply by changing the value of the axion ${\rm Im}\ \rho$, as the axion appears nowhere else in the potential.

Given this change in the axion minimum, the potential for all
the other fields on the tip of the throat will look
the same for the $\D3$-brane as it did for the D3-brane, and the volume and brane
locations will have the same solutions as they did for the
D3-brane case, equivalent to our previous analysis.

In summary, the change from a D3-brane at the bottom of the throat
to a $\D3$-brane modifies only the stabilized value of the axion,
while preserving the overall volume and brane vacua.  Moduli
stabilization forces restrict D3-branes and $\D3$-branes to the
same subloci of the three-sphere.

\section{Enumerating brane forces and angular inflation}
\label{InflationSec}
\setcounter{equation}{0}

In the preceding sections we have seen that D3-branes and
$\D3$-branes are driven to the same loci by moduli stabilization
forces. Moreover, these loci often
include flat directions protected by isometries.  A natural question is whether inflation
could occur as a D3-brane moves along such a trough in the angular directions ({\it{cf.}}
\cite{Kallosh}).

To examine this question, we need to consider {\em all} the possible forces acting on the branes, and to estimate their relative strengths and effects.  Hence, we will now compute the primary
contributions to the potential for the scalar degrees of freedom of
a brane and an antibrane in a
KS throat.  The enumeration and comparison of all forces on the branes is also interesting in its own right. These contributions are: the brane-antibrane tachyon potential, the brane-antibrane Coulomb
potential, the potential from bulk breaking of isometries, and the potential from moduli stabilization.

Equipped with this information, we will then consider, in \S\ref{AngInflateSec}, a possible inflationary scenario
involving brane-antibrane
attraction along the angular directions of the $S^3$ at the tip.
We find that this scenario does not produce a sufficient number of e-folds. Our results apply directly only to a
KS throat, and the difficulties may be more or less severe in other throat constructions.

\subsection{The brane/antibrane potential}
\label{BranePotSec}

\subsubsection*{Brane-antibrane interactions}

A brane and an antibrane interact at long distances by their mutual
Coulomb interaction and at short distances by the appearance of an
open string tachyon. The Coulomb interaction in a warped throat can
found by computing the perturbation of the warp
factor/five-form field $\Phi_+$
by a D3-brane in the warped space,
which is then used to compute the energy of a $\D3$-brane
as a function of the D3-brane position. The details of the
calculation are given in Appendix B; for small angular and radial
separations, the potential takes the form,
\begin{equation}
V_{D \overline{D}} = 2 \,  T_{D3} \, a_0^4 \left( 1 - \frac{4 \pi g_s {\alpha^{\prime}}^2}{R_{S^3}^4}\frac{1}{(\rho^2 + (\Delta\Omega)^2)^2} + \ldots \right) \,,
\label{eq:coulomb}
\end{equation}
where $\rho\equiv |\vec\tau-\vec\tau_1|$ and $\Delta\Omega$ are the dimensionless coordinate radial separation
and $S^3$ angular separation between the brane and antibrane on the approximate $\RR\times S^3$
geometry near the tip, respectively.  (In this section we will keep explicit factors of
$\alpha^{\prime}$ for clarity, though we have set $\alpha^{\prime} \to 1$ elsewhere in the paper.)

The brane-antibrane tachyon can begin to condense when the proper
(warped) transverse
distance between a D3-brane and $\D3$-brane is
of order a string length.  Near the tip we see from the metric (\ref{TipMetric})
that proper warped separations of
order a string length correspond to dimensionless coordinate distances
\begin{eqnarray}
\label{TachyonRad}
\rho, \Delta\Omega \sim (g_s M)^{-1/2} \,.
\end{eqnarray}
However, we require $g_s M \gg 1$ to be in the supergravity limit, so order-one coordinate distances are far outside the range of the tachyon; only at the parametrically smaller separation
(\ref{TachyonRad}) does the tachyon set in.  This justifies focusing on the Coulomb potential for most of the evolution of a brane-antibrane pair at the tip.

\subsubsection*{Approximate isometries and bulk effects}

The
KS throat has the exact isometry $SO(4)\simeq
SU(2)\times SU(2)$ which is enlarged to $SU(2)\times SU(2)\times
U(1)$ (the symmetry group of $T^{1,1}$, the base of the cone) away
from the tip of the throat.  These geometric isometries are reflected in the form of the K\"ahler potential
(\ref{KahlerPot}), which depends only on the radial coordinate \cite{Candelas}:
\begin{equation}
\label{LongKahlerEqn}
k(z_i,\bar{z}_i) = 2^{-1/2}\epsilon^{4/3} \int (\sinh 2\tau-2\tau)^{1/3} \sim
	\cases{\frac{3}{2}\left(\sum_i|z_i|^2\right)^{2/3} = r^2 & $r\gg \epsilon$ \cr
			k_0 + 18^{-1/6} \epsilon^{-2/3}\left(r^3-\epsilon^2\right)
				 & $r\rightarrow \epsilon$.\cr}
\end{equation}
Away from the tip, the K\"ahler potential does
not depend on any of the angular coordinates of $T^{1,1}$, and at
the tip it does not depend on the coordinates on the
$S^{3}$.\footnote{ Let us note that the presence of isometries is
not generic, and one could expect a `typical' throat to have no
isometries at all.}

K\"ahler potential isometries of this sort are important because
they strongly restrict the form of the potential felt by a probe
brane. In particular, if a particular isometry of the K\"ahler
potential is also respected by the superpotential
-- determined for us by the embedding of the wrapped D7-branes --
 then the total
potential will be flat along the associated direction, as we saw repeatedly for the
moduli spaces in
\S\ref{ConifoldVacSec}.

However, the throat is attached to a larger, compact geometry, which in general need not preserve any of these isometries.  In fact, a compact
Calabi-Yau cannot have exact continuous isometries, so when a finite
throat is glued into a compact bulk, the throat isometries must be
broken, perhaps weakly, by bulk effects.

This breaking may be
described as arising from vevs of certain operators of the dual
gauge theory \cite{Ofer}. An operator of dimension $\Delta$ which
breaks the isometry in the UV gives rise to a mass in the IR,
\begin{equation}
\label{BulkMass}
m_{bulk}^2 \sim (g_s M\alpha')^{-1} a_0^{\Delta-2}.
\end{equation}
Provided the operators in question have sufficiently large
dimensions, these symmetry-breaking effects are suppressed by powers
of the warp factor.  Perturbations associated to relevant
operators $\Delta < 4$ would correspond to completely changing the KS throat geometry and would violate our assumption that
the compactification contains a KS throat, while those associated to marginal operators $\Delta = 4$ correspond to shifts in moduli that we have taken to be fixed.
Thus, the bulk effects of interest are those associated to irrelevant operators with $\Delta > 4$.  The corresponding perturbations are suppressed by powers of the warp factor.

We are assuming that the bulk geometry before moduli stabilization is of the GKP type
even outside the throat, obeying (\ref{ISD}) as described by \cite{GKP}.  Consequently the perturbations from bulk effects (\ref{BulkMass}) can produce no potential for a D3-brane, which feels a flat potential in any
such background.  A $\D3$-brane, on the other hand, will in general be sensitive to bulk effects.  Hence we may think of the warping as protecting the $\D3$-brane potential from corrections by producing a sequestered region
that preserves some approximate isometries.

\subsubsection*{Mass terms from D7-branes}

As we have seen in the preceding sections, the wrapped
branes that generate a nonperturbative superpotential for the
K\"ahler moduli also generate a potential for D3-brane motion,
through the correction computed in \cite{BHK,Baumann}.
However, the form of this potential depends on the precise embedding
of the wrapped branes, and there exist simple embeddings that
preserve some of the isometries of $T^{1,1}$; for example, the
Karch-Katz embedding (\ref{eq:KarchKatzEmbed})
preserves a full
$SO(2) \times SO(2)$ isometry with possible $S^1$ moduli spaces.  If we consider such an embedding, a D3-brane and a $\D3$-brane will be restricted to the same locus
on the tip by the nonperturbative forces; however, the mutual Coulomb attraction will be present, and in general the $\D3$-brane will also feel the bulk effects forces we have discussed.
We note that for embeddings which do not admit supersymmetric vacua on the tip (e.g., the
Ouyang embedding), the potential separating the D3 from the $\D3$ which is drawn to the tip by the warped background may also interfere with inflation.

The simplest brane-antibrane inflation
model is one in which the D3-brane and the $\D3$-brane are separated along a flat direction and feel only the Coulomb attraction.  Could the locus on the tip selected by the moduli-stabilization forces include such a flat direction?  To address this, we first
need to know whether the moduli stabilization forces are stronger than the bulk effects.  This is
important: if the bulk isometry-breaking effects gave a leading
contribution to the potential, then the $\D3$-brane could be separated from the
D3-brane by a potential barrier that is large compared to the
potential from moduli stabilization, and also large compared to the
Coulomb interaction.  If instead the moduli stabilization effects are dominant, they will restrict the D3-brane and $\D3$-brane to a common moduli space.  Then, the weaker bulk effects will draw the $\D3$-brane to some point in this space, but will not affect the D3-brane.  In the end, the D3-brane is free to move under the influence of the Coulomb force alone, and brane inflation can proceed.  We will now show that the moduli-stabilizing force is generically the dominant one.

The angular potential induced by the D7-branes is somewhat
complicated, but we only need a parametric estimate that includes
powers of the warp factor.  For example, there is a term
\beq \label{sampleterm}
V_{1} = \left|\hat{f}(Y)\right|^2 A_0^2 \frac{a^2(\rho + \bar\rho)
e^{-(\rho + \bar\rho)}}{3 e^{8u}} \equiv V_0 \left|\hat{f}(Y)\right|^2 \,. \eeq
that only depends on the angles via the factor $|\hat{f}(Y)|^2 \equiv |\mu^{-1}f(Y)|^2$, as $V_0$ depends only on the radial coordinate.  Here $\mu$ is the constant term appearing in $f$, {\it{e.g.}} in the simple Kuperstein embedding $f=z^A - \mu$ or the Ouyang embedding $f=w_i-\mu$.  The precise functional form of $f$ is not, however, important at present.

We will consider terms of this form to get an estimate of the relative strengths of moduli stabilization forces and bulk effects on the $S^3$.  Any other terms from moduli stabilization that are weaker than (\ref{sampleterm}) can be neglected in comparison, while if any other terms are stronger they provide an even stronger potential.  Thus, by showing that (\ref{sampleterm}) is stronger than the bulk effects, we can establish our hypothesis.

Before proceeding, we note that requiring that a $\D3$-brane at the tip can lift this configuration to a metastable de Sitter vacuum implies that $V_0 \propto T_{D3}a_0^4$ \cite{KKLT}.  This is because the energy of such a $\D3$-brane is proportional to $T_{D3}a_0^4$, while the negative cosmological constant associated to moduli stabilization is proportional to $-V_0$.  If $V_0 \gg T_{D3}a_0^4$, the net cosmological constant is negative.  On the other hand, if $V_0 \ll T_{D3} a_0^4$, the potential is dominated by the $\D3$-brane, which drives a runaway decompactification.  Only when these two effects are balanced, $V_0 \sim T_{D3}a_0^4$, does a metastable de Sitter vacuum arise.

For simplicity we focus on a particular direction $\theta$ on the $S^3$.  The canonically-normalized field associated to $\theta$ is
\beq \label{CanonVar} \vartheta \equiv c a_0 \theta \eeq
where $c = (T_{D3} g_s M \alpha^{\prime})^{1/2}e^{-2u}$.  (In the remainder of this discussion we will omit the factor $e^{2u}$ and focus on the more important dimensionful quantities and powers of the warp factor.)  A key consequence is that the curvature of the potential in the canonical $\vartheta$ direction is parametrically greater than the curvature in the $\theta$ direction: \beq \frac{\partial^2 V_{1}}{\partial \vartheta^2} = \frac{1}{c^2 a_0^{2}} \frac{\partial^2 V_{1}}{\partial \theta^2} \, . \eeq  Combining this result with the condition $V_0 \sim T_{D3}a_0^4$, we conclude that moduli stabilization gives rise to a mass-squared in the canonical $\vartheta$  direction of order \beq m^2_{D7} = \frac{\partial^2 V_{1}}{\partial \vartheta^2} \sim \frac{a_0^2}{g_s M \alpha^{\prime}} \frac{\partial^2}{\partial \theta^2}\left|\hat{f}(Y)\right|^2 \, . \eeq  On the other hand, the typical mass-squared from bulk effects is $m^{2}_{bulk} \sim ({g_s M \alpha^{\prime}})^{-1} a_0^{3.29}$.  We conclude that $m^{2}_{D7}$ will be much larger than $m^2_{bulk}$, provided that $(\partial^2 \left|\hat{f}(Y)\right|^2 /\partial\theta^2) \gg a_0^{1.29}$, {\it{i.e.}} unless the dimensionless curvature in the $\theta$ direction is parametrically small.
However, there is no reason to expect this curvature to be very small unless the parameters characterizing $\hat{f}$ are tuned to take extreme values; generic ratios like $\epsilon/\mu$ of order unity thus lead to the generic case of the moduli stabilization forces being hierarchically larger than the forces from bulk effects.

We will now illustrate this in a simple example, the Kuperstein embedding with $n=1$.  In this case $\hat{f}$ has the form
\beq \hat{f} = 1 -\frac{z^1}{\mu}  \,. \label{KuperA} \eeq
The term (\ref{sampleterm}) can then be written
\beq \label{newsampleterm}
V_{1} = V_0 \left|1-\frac{z^1}{\mu}\right|^2 \, , \eeq
where the constant $V_0$ is independent of the angles.
The vacuum is a point, sitting at $z^2=z^3=z^4 = 0$, $z^1 = \epsilon$.
Consider the special line $\psi=\pi$, $\phi=0$,
with $\theta$ allowed to vary to fill out the line.  Along this line, \beq z^1 = \epsilon ~ {\rm{sin}}\frac{\theta}{2} \,, \eeq
({\it{cf.}} Appendix A) and so $z^1$ ranges between $\epsilon$
(the vacuum)
and $0$.

The curvature of the potential in the $\vartheta$ direction, to leading order in $\epsilon/\mu$, is \beq
\frac{\partial^2 V_{1}}{\partial \vartheta^2}
\sim \frac{a_0^2}{g_s M \alpha^{\prime}}~ \frac{\epsilon}{\mu}{\rm{sin}}\frac{\theta}{2} \, . \eeq
Hence, the mass-squared in the canonical $\vartheta$ direction is of order \beq m_{D7}^2 \sim \frac{a_0^2}{g_s M \alpha^{\prime}}~\frac{\epsilon}{\mu} \, .\eeq
We conclude that $m^{2}_{D7}$ can be much larger than $m^2_{bulk}$, provided that
\beq \label{muepsiloncondition} \frac{\epsilon}{\mu} \gg a_0^{1.29} \, . \eeq  The parameter $\mu$ measures the minimal radial location reached by the D7-brane, while $\epsilon$ characterizes the deformation of the conifold.  Thus, the condition (\ref{muepsiloncondition}) is just the requirement that the D7-brane reaches sufficiently far into the throat, which is
generically satisfied.
In conclusion, it appears that the mass terms from moduli stabilization
are typically
large compared to those from bulk effects.

\subsection{Angular inflation}
\label{AngInflateSec}

Brane-antibrane inflation is attractive as an inflationary scenario
because the Coulomb interaction is weak at long distances, leading
to a flat potential \cite{DvaliTye}.  Additional corrections to the
potential coming from the conformal coupling of the inflaton to the
background and from moduli stabilization ruin the flatness of the
potential \cite{KKLMMT}, but in some fine-tuned setups these
contributions might possibly be tuned to cancel each other \cite{etaProblem,Uplifting,BaumannToAppear}.

In most brane inflation
scenarios the angular directions are ignored and inflation proceeds along the radial coordinate.  (For a study of antibrane-only inflation in the angular directions see \cite{Giant}.  The mirage cosmology arising from angular motion in a warped throat was considered in \cite{Gregory}.)  However,
because the angular directions are protected by symmetries, the potential along these directions could be flat enough for inflation
to take place.  Moreover, fluctuations in such flat angular directions could give important corrections to scenarios based on
radial motion.

In light of this, we will consider the possibility of angular inflation -- relative angular motion of a brane/antibrane pair living in a shared flat direction
at the bottom of the throat.  The $\D3$-brane may feel forces from isometry-breaking bulk effects, but assuming these are weaker than the moduli stabilization forces -- as we have shown is generically the case -- these will merely pull the $\D3$-brane to a point on the shared flat direction.

The D3-brane then feels nothing but the Coulomb potential, and we
now determine whether this potential is flat enough for inflation.  Without loss of generality, we focus on motion in the $\theta$ direction.
Rewriting (\ref{eq:coulomb}) in terms of the canonically normalized field $\vartheta$ (\ref{CanonVar}), we find
\begin{equation}
V_{D \overline{D}} = 2 \,  T_{D3} \, a_0^4 \left( 1 - \frac{T_{D3} \, a_0^4 \, e^{-8u}}{2\pi^2\vartheta^4} + \ldots \right) \,,
\end{equation}
where we also used (\ref{a0def}).
We now compute the slow-roll parameter $\eta \equiv M_p^2 V''/V$,
where the derivatives are with respect to the canonical variable $\vartheta$; we find
\begin{eqnarray}
\eta = - M_p^2 \, \frac{10}{\pi^2}\frac{T_{D3}\, a_0^4 \,e^{-8u}}{\vartheta^6}\,.
\end{eqnarray}
Naively this seems to be enormously suppressed by the exponential warp factor $a_0^4$.

However,
the situation is more complicated.
The canonically normalized variable $\vartheta$,
as is apparent in its definition (\ref{CanonVar}), is
compressed by the warping to a field space of exponentially small size.
The maximum value of $\vartheta$ is not some order-one quantity like $\pi$, but rather
\begin{eqnarray}
\vartheta_{max} \sim a_0\, (T_{D3} g_s M \alpha^{\prime})^{1/2}\, e^{-2u} \,.
\end{eqnarray}
The minimum value of $|\eta|$ over the field space of $\vartheta$ occurs at $\vartheta_{max}$; we find
\begin{eqnarray}
\label{etamin}
|\eta_{min}| \sim \frac{10 M_p^2 e^{4u}}{T_{D3}^2\, a_0^2 \, \pi^2 (g_s M \alpha^{\prime})^3} \,.
\end{eqnarray}
This is exponentially large; the potential is not suitable for inflation.

What we have learned is
that the warp factor tends to flatten out the potential, which helps inflation; however, it also
compresses the field space in the canonical variable, which hurts.  We find that the latter effect wins, and there is too little field space to permit brane/antibrane inflation in the angular directions.
One may also see this with the number of $e$-folds:
\begin{eqnarray}
N_e &=& \frac{1}{M_p^2}\int \frac{V}{V'}d\vartheta = \frac{\pi^2}{10}\, \frac{T_{D3}^2}{M_p^2}\, a_0^2 \, (g_s M \alpha^{\prime})^3 e^{-4u} = {1 \over \eta} \,.
\end{eqnarray}
Hence, over the available field space the number of $e$-foldings is exponentially small.

We may
inquire whether parametric factors in (\ref{etamin}), other than the warping, may ameliorate the situation.  Recall that $M_p^2 = \kappa_{10}^{-2} V_6$ where
$\kappa_{10}^{2}=\frac{1}{2}(2\pi)^7 g_s^2 {\alpha^{\prime}}^4$, and
the six-dimensional volume is conservatively bounded by $V_6\geq
R_{AdS}^6$.
Using the relations
\begin{eqnarray}
R_{AdS}^4 =  \frac{27\pi}{4}g_s K M {\alpha^{\prime}}^2 \,, \quad a_0 \sim  e^{-2\pi K/(3g_s M)} \,,
\end{eqnarray}
we indeed find that
\beq N_{e} \ll 1 \,. \eeq
We conclude that motion along the angular directions gives rise to a negligible amount of inflation.
In particular, this implies that for scenarios involving radial motion of a D3-brane, it is not necessary to worry about
contributions from angular motion: the angular directions are not flat enough to have an important effect.

\section{Conclusions}
\label{ConclusionsSec}
\setcounter{equation}{0}

Although D3-branes enjoy a no-force condition in no-scale flux
compactifications, this property is lost in the presence of the
nonperturbative effects that can stabilize the K\"ahler moduli.  A
D3-brane at a generic point in such a compactification is
non-supersymmetric,
and will generally be driven to a vacuum
on which supersymmetry is restored.

We studied the general equations for supersymmetric vacua in a moduli-stabilized
compactification, and then enumerated supersymmetric vacua in the explicit example of
the $S^3$ at the tip of the warped deformed conifold,
for various
configurations of the wrapped D7-branes that generate the
nonperturbative superpotential.  We found examples in which the
D3-brane vacua had real dimension two, one, and zero, and we
argued that the last of these is the generic result in a compact
Calabi-Yau.  We also demonstrated that $\D3$-branes are confined
by nonperturbative forces to the same loci as the D3-branes,
preserving the usual exit from brane/antibrane inflation.

Finally, we asked whether the flat angular directions associated
with continuous D3-brane moduli spaces could be relevant in
D-brane inflation, as  a D3-brane moving along such an angular
direction under the influence of a weak  Coulomb force might be
expected to give rise to inflation.  We showed that this is not
possible because
-- despite the naive flattening of the potential by the warp factor --
the canonical field distance along the angular
directions becomes exponentially small.  Thus, angular motion of a
D3-brane does not give rise to prolonged inflation.

\section*{Acknowledgments}
It is a pleasure to thank Daniel Baumann, Shanta de Alwis, Anatoly Dymarsky, Min-xin
Huang, Thomas Grimm, Shamit Kachru, Ben Shlaer, and Herman Verlinde for helpful
discussions.
L.M. thanks the KITP Santa Barbara and the theory groups at Stanford University, the University of Texas, and the University of Michigan
for hospitality during the completion of this work.  The work of O.~D.\ was supported by DOE grant DE-FG02-91-ER-40672.  The work of L.~M.\ was supported in part by DOE grant
DE-FG02-90ER-40542.  The work of G.~S.\ and B.~U.\ was supported in part by
NSF CAREER Award No. PHY-0348093, DOE grant DE-FG-02-95ER40896, a
Research Innovation Award and a Cottrell Scholar Award from Research
Corporation.

\appendix

\section{Coordinates on the deformed conifold}
\setcounter{equation}{0}

Here we collect a few facts concerning the various coordinates
parameterizing the deformed conifold.  It is defined via the
equation
\begin{eqnarray}
\sum_{A=1}^4 (z^A)^2 =-2 (w_1w_2 - w_3 w_4) = \epsilon^2 \,,
\end{eqnarray}
and the D7-brane embeddings we use are given in terms of one or
the other of these sets of coordinates.  These coordinates can be
related to coordinates on the $S^3$ at the bottom of the throat as
follows.
(We follow \cite{Candelas} with some modifications to
their notation.)  Define the matrix $W$ as
\begin{eqnarray}
W \equiv L W_0 R^\dagger \,, \quad \quad W_0 \equiv
\pmatrix{\epsilon/\sqrt{2} \; \sqrt{r^3-\epsilon^2} \cr 0 \;\;\;\;
-\epsilon/\sqrt{2}} \,,
\end{eqnarray}
where $L,R$ are $SU(2)$ matrices parameterized by three Euler
angles each.  (We are using the standard $r$-variable on the
conifold, related to that in \cite{Candelas} by
$r = (r_{there})^{2/3}$.)  We choose the following convention,
\begin{eqnarray}
\label{Wdef} W = \pmatrix{-w_3 \; w_2 \cr -w_1 \; w_4} = - {1 \over
\sqrt{2}} \pmatrix{z^3 + i z^4 \;\;\;\; z^1 - i z^2 \cr z^1 + i
z^2 \;\; -z^3 + i z^4} \,,
\end{eqnarray}
where we have chosen the $w$'s so as to agree with (32)-(35) of
\cite{Baumann} when we use the  parameterization of Euler angles
given in (2.24)-(2.25) of \cite{Candelas}.  One indeed finds that
\begin{eqnarray}
{\rm det}\ W = w_1 w_2 - w_3 w_4 = -{1\over 2} \sum_{A=1}^4
(z^A)^2 =  - {1 \over 2} \epsilon^2 \,,
\end{eqnarray}
as required.   At generic $r > \epsilon^{2/3}$, one of the six
Euler angles in $L$ and $R$ is redundant, and the remaining five
along with $r$ parameterize the deformed conifold.  For $ r \gg
\epsilon^{2/3}$ the deformed conifold is well-approximated by the
singular conifold, with the angles parameterizing $T^{1,1}$ and
with the metric (\ref{ConifoldMetric}).

The $S^3$ is at $r=\epsilon^{2/3}$, where $W_0(r = \epsilon^{2/3})
\propto \sigma_3$.  At this point three of the six angles become
redundant.  The matrix $T \equiv -  {\sqrt{2} \over \epsilon} W
\sigma_3$ turns out to be an element of $SU(2)$ (the minus sign is
just for convenience), and its three independent angles can be
used to parameterize the $S^3$.  Let the parameterization be
\begin{eqnarray}
\label{ConifoldSphereCoords}
T \equiv {\sqrt{2} \over \epsilon} \pmatrix{w_3 \;w_2 \cr w_1 \;
w_4} = \pmatrix{e^{i (\psi+\phi)/2} \cos \theta/2 \;\;\;\;  i
e^{i (\psi-\phi)/2} \sin \theta/2 \cr i e^{-i (\psi-\phi)/2} \sin
\theta/2 \;\;\;\; e^{-i (\psi+\phi)/2} \cos \theta/2} \,,
\end{eqnarray}
which is the standard Euler angle presentation of angles $\{ \psi,
\theta, \phi \}$ on $S^3$ with the associated metric
(\ref{3Sphere}).  (These angles are implicitly related to those of
the approximate $T^{1,1}$ at $r \gg \epsilon^{2/3}$, but we will
not need to work out this relation.)  We note that on the $S^3$,
the $w_i$ obey nontrivial relations:
\begin{eqnarray}
w_1 = - \overline{w}_2 \,, \quad \quad w_3 = \overline{w}_4 \,.
\end{eqnarray}
One may furthermore show using (\ref{Wdef}) that the $S^3$ angles
are related to the $z^A$ by
\begin{eqnarray}
\label{zCoords}
z^1 &=& \epsilon \sin {\theta \over 2} \sin {\psi - \phi \over
2} \,, \quad
z^2 = \epsilon \sin {\theta \over 2} \cos {\psi - \phi \over 2} \,, \\
z^3 &=&  \epsilon \cos {\theta \over 2} \cos {\psi + \phi \over
2} \,, \quad z^4 =  \epsilon \cos {\theta \over 2} \sin {\psi +
\phi \over 2} \,. \nonumber
\end{eqnarray}
We see that in this case, the $S^3$ is a real slice of each $z^A$
coordinate.

\section{${\rm D3}$-$\D3$ potential in the deformed conifold}
\setcounter{equation}{0}

The potential between an antibrane located at the tip of the
deformed conifold (KS throat) and a brane located at a different
coordinate value in the throat can be found by calculating the
energy of the $\D3$-brane in a background geometry perturbed
by a D3-brane.

In this Appendix we compute the potential between a $\D3$-brane at the tip of the deformed conifold and and a D3-brane at or near the tip by considering the leading backreaction of the D3-brane on the background.  The metric in this region, $\tau \approx 0$, is approximately (\ref{TipMetric}), which is simply that of $\RR^3\times S^3$ up to an overall factor of $\epsilon^{4/3}$; we use $\vec\tau$ to denote a vector $(\tau, \Omega_2)$ on $\RR^3$ and $\Omega$ for the three coordinates on $S^3$.

Adding a D3-brane at position $\vec{y}_1 = (\vec\tau_1,\Omega_1)$
on the $\RR^3\times S^3$ preserves the imaginary self-dual conditions (\ref{ISD}), and modifies $\Phi_+$ to
\begin{eqnarray}
\label{PhiExpand}
\Phi_+^{-1} =  (\Phi_+^{-1})_0 + \phi_+^{-1} \,,
\end{eqnarray}
where $(\Phi_+^{-1})_0$ is the background value (\ref{WarpFactor}), and following equation (\ref{PhiEqns}) $\phi_+^{-1}$ solves
\begin{eqnarray}
\label{PertEqn}
- \tilde\nabla^2 \, \phi_+^{-1} =  8 \pi^4 g_s \, {\delta^6(y - y_1) \over \sqrt{\tilde{g}_6}}\,.
\end{eqnarray}
We define $g^0_{mn} \equiv \epsilon^{-4/3} \tilde{g}_{mn}$ as the standard metric on $\RR^3 \times S^3$,
\begin{eqnarray}
g^0_{mn} dy^m dy^n = d\vec\tau^2 + d \Omega_3^2 \,,
\end{eqnarray}
in terms of which (\ref{PertEqn}) becomes
\begin{eqnarray}
- \nabla_0^2 \, \phi_+^{-1}    = {8 \pi^4 g_s  \over \epsilon^{8/3}} \, {\delta^6(y - y_1) \over \sqrt{g^0_6}}  \equiv {\cal C} \, {\delta^6(y - y_1) \over \sqrt{g^0_6}} \,.
\end{eqnarray}
We thus see that $1/({\cal C} \phi_+)$ is a Green's function $G$ on $\RR^3 \times S^3$,
\begin{equation}
\nabla_0^2\,  G(|\vec\tau-\vec\tau_1|,\Omega-\Omega_1) =
-\frac{\delta^3(\vec\tau-\vec\tau_1)\delta^3(\Omega-\Omega_1)}{\sqrt{g^0_6}} \,.
\label{LaplaceS3}
\end{equation}
Since the Green's function should only depend on the separation
between the coordinates, we will take the following separable
ansatz with $\rho\equiv |\vec\tau-\vec\tau_1|$ and $\Delta \Omega
\equiv \Omega-\Omega_1$,
\begin{equation}
G(\rho,\Delta\Omega) = \sum_{L} A_L(\Omega_1) g_{L}(\rho)
Y_{\{L\}}(\Omega) \label{eq:GreenExpansion}
\end{equation}
where the functions satisfy the following differential equations
away from $\rho, \Delta\Omega = 0$,
\begin{eqnarray}
\label{eq:LaplaceR3}
\nabla^2_{\RR^3} g_L(\rho) + k_L^2 g_L(\rho) &=& 0 \\
\nabla^2_{S^3} Y_{\{L\}}(\Omega) + \frac{L(L+2)}{R_{S^3}^2}
Y_{\{L\}}(\Omega) &=& 0, \label{eq:LaplaceS3}
\end{eqnarray}
with $k_L^2 = -L(L+2)$, $L\in \ZZ$.  The
solutions to the three-sphere Laplace equation
Eq.(\ref{LaplaceS3}) are hyperspherical harmonics, which have the
following useful properties \cite{Hypergeometric},
\begin{eqnarray}
\int Y_{\{L\}}^*(\Omega) Y_{\{L\}}(\Omega) d\Omega &=& \delta_{\{L\},\{L'\}} \\
\sum_{\{L\}} Y_{\{L\}}^*(\Omega_1) Y_{\{L\}}(\Omega_2) &=& \delta^{(D-1)}(\Omega_1-\Omega_2) \\
\sum_{[L]} Y_{\{L\}}^*(\Omega_1) Y_{\{L\}}(\Omega_2) &=&
\frac{2{\mathcal
L}+1}{4\pi^{D/2}}\Gamma(D/2-1)C_L^{D/2-1}(\cos\alpha)
\label{eq:hypersphereAddition}
\end{eqnarray}
where the first line is the orthogonality of the functions, the
second line is completeness, and the third line is the generalized
addition theorem for hyperspherical harmonics with $\alpha$ the
angle between the two vectors with angles $\Omega_1$ and
$\Omega_2$ on the three-sphere, ${\mathcal L} = L+(D-3)/2$.  By
$\{L\}$ we mean the set of all angular quantum numbers, by $[L]$
we mean only secondary angular quantum numbers, and by $L$ we mean
the primary angular quantum number appearing in the differential
equation.  For example, for $D=3$ we have $\{L\} = \{\ell,m\}$,
$[L] = m$ and $L=\ell$. The functions $C_n^{\alpha}$ are
Gegenbauer polynomials.

The $\RR^3$ differential equation Eq.(\ref{eq:LaplaceR3}) is the
well-known Helmholtz equation, with Green's function solution
\begin{equation}
g_L(\rho) = \frac{e^{i k_L \rho}}{4 \pi \rho}.
\end{equation}
The coefficients of the expansion Eq.(\ref{eq:GreenExpansion}) can
be found by integrating across the angular part of the delta
function argument, which gives the full solution,
\begin{equation}
G(\rho,\Delta\Omega) = \sum_{\{L\}} \frac{e^{i k_L \rho}}{4\pi\rho
} Y_{\{L\}}(\Omega) Y_{\{L\}}^*(\Omega_1)
    = \frac{1}{16\pi^3 \rho \sin(\Delta\Omega)}\sum_L (2{\mathcal L}+1)e^{i k_L \rho} \sin\left((L+1) \Delta\Omega\right) \,,
\label{eq:ProductGreenFcn}
\end{equation}
where the addition theorem Eq.(\ref{eq:hypersphereAddition}) was used to
make the dependence on $\Delta\Omega$ explicit.  Note that
\begin{equation}
C_L^1(\cos(\Delta\Omega)) =
\frac{\sin((L+1)\Delta\Omega)}{\sin(\Delta\Omega)}.
\end{equation}
For small $\Delta\Omega$, only the large $L$ terms in the sum
contribute, so we can take the approximation $k_L =
i\sqrt{L(L+2)} \approx i L$. The sum
Eq.(\ref{eq:ProductGreenFcn}) can now be done in closed form and
becomes
\begin{equation}
G(\rho,\Delta\Omega) = \frac{\left(e^{2\rho}-1\right)/\rho}{32\pi^3 \left(\cosh\rho-\cos(\Delta\Omega)\right)^2}\,. \label{eq:GreenFull}
\end{equation}
It is straightforward to see that Eq.(\ref{eq:GreenFull}) reduces
to the flat space limit for $\rho\ll 1$, $\Delta\Omega
\ll 1$,
\begin{equation}
G(\rho,\Delta\Omega) \approx \frac{1}{4\pi^3 (\rho^2+\Delta\Omega^2)^2} \,. \label{eq:GreenFlat}
\end{equation}
Thus we obtain the perturbation $\phi_+^{-1} = {\cal C} G$.  The potential for the brane/antibrane system is then $T_{D3} \Phi_+$ (\ref{BraneCoupling}) with $\Phi_+$ given by (\ref{PhiExpand}), which becomes to leading order at the tip,
\begin{eqnarray}
V_{D \overline{D}} = 2 \,  T_{D3} \, a_0^4 \left( 1 - {4 \pi g_s \over R_{S^3}^4} {1 \over (\rho^2 + (\Delta\Omega)^2)^2} + \ldots \right) \,,
\end{eqnarray}
with $R_{S^3}^2 \equiv g_s M$.

\section{Off-tip SUSY vacua for Ouyang embeddings}
\setcounter{equation}{0}

We saw in \S\ref{ACRSec} that the Ouyang type of embeddings
\begin{equation}
f = w_i -\mu
\label{OuyangEmbedApp}
\end{equation}
do not admit supersymmetric vacua for D3-branes at the tip of
the deformed conifold.  There generically are, however, supersymmetric vacua off the tip, as we now demonstrate.

We keep the gauge choice (\ref{VarElim}) of eliminating $w_1$, and consider the Ouyang embedding (\ref{OuyangEmbedApp}) with $i=2$.
Clearly, $\partial_{w_3} f = \partial_{w_4} f = 0$,
so we must have
\begin{eqnarray}
\partial_3 \left(\zeta + \frac{a}{3} k\right) &=& \frac{a}{3}\partial_{r^3} k \left(\frac{w_4}{w_2}\overline{w}_1+\overline{w}_3\right) = 0 \\
\partial_4 \left(\zeta + \frac{a}{3} k\right) &=& \frac{a}{3}\partial_{r^3} k \left(\frac{w_3}{w_2}\overline{w}_1+\overline{w}_4\right) = 0\, .
\end{eqnarray}
First we note that  $\partial_{r^3} k \neq 0$ in general (\ref{LongKahlerEqn}).  One way of satisfying these relations is with $|w_1|^2 = |w_2|^2$ and $\overline{w}_3 = -w_4\overline{w}_1/w_2$; this corresponds to the tip.  However, they may also be solved by
\begin{eqnarray}
w_3 = w_4 = 0 \,.
\label{NewCond}
\end{eqnarray}
This locus intersects the tip, but extends off it; it is one-complex-dimensional, parameterized by $w_2 \neq 0$, while its intersection with the tip is $|w_2|^2 = \epsilon^2/2$, which is one-real-dimensional.

Choosing the constraint (\ref{NewCond}) allows us to satisfy $\partial_{w_3} k = \partial_{w_4} k =0$ without imposing $\partial_{w_2} k = 0$, which would lead to no solution.  The final vacuum equation is
\begin{eqnarray}
\partial_{w_2} \left(\zeta + \frac{a}{3}k\right) &=& -\frac{1}{n(w_2-\mu)} +
\frac{a}{3}\partial_{r^3}k \left(- \frac{w_1}{w_2}\overline{w}_1+\overline{w}_2\right) = 0 \,.
\end{eqnarray}
We may use (\ref{NewCond}) and the conifold equation, which becomes $w_1 = -\epsilon^2/(2 w_2)$, to obtain
\begin{eqnarray}
{w_2 \over w_2 - \mu} = F(|w_2|^2) \left( |w_2|^2 - {\epsilon^4 \over 4 |w_2|^2} \right) \,.
\label{OffTipEqn1}
\end{eqnarray}
Here $F(|w_2|^2)$ is a {\em real} function, consisting of real coefficients times $\partial_{r^3} k$, which is evaluated as a function of $|w_2|^2$.  Thus since the right-hand-side is real, we must require $\mu/w_2$ to be real as well.  Defining $\mu \equiv m e^{i \chi}$  with $m>0$, and $w_2 \equiv w e^{i \delta}$, we impose this by requiring $\delta=\chi$, while allowing the real $w$ to take either sign.
Hence (\ref{OffTipEqn1}) becomes
\begin{eqnarray}
4 w^3 = (w-m) F(w^2) (4 w^4 - \epsilon^4) \,.
\end{eqnarray}
This equation is in general a polynomial of odd order in $w$; consequently we expect at least one real solution for $w$, though in general there may be discretely many.
Hence we indeed expect to find off-tip vacua for the Ouyang embedding; the solutions would only lie on the tip for the special value $w = \epsilon/\sqrt{2}$, which one can see by inspection is not a solution since the right-hand-side vanishes, consistent with the analysis of \S\ref{ACRSec}.

\end{document}